# AutoSAS: a new human-aside-the-loop paradigm for automated SAS fitting for high throughput and autonomous experimentation

Duncan R. Sutherland[1,2], Rachel Ford[3], Yun Liu[3,4], Tyler B. Martin[1*], Peter A. Beaucage[3*]

[1]Materials Science and Engineering Division, National Institute of Standards and Technology, Gaithersburg, MD 20899, United States
[2]Department of Physics, University of Colorado at Boulder, Boulder, CO 80309, United States
[3]NIST Center for Neutron Research, National Institute of Standards and Technology, Gaithersburg, MD 20899, United States
[4]Department of Chemical Engineering, University of Delaware, Newark, DE 19716, United States
* to whom correspondence should be addressed, TBM: tyler.martin@nist.gov, PAB: peter.beaucage@nist.gov

**Abstract**

The advancement of artificial-intelligence driven autonomous experiments demands physics-based modeling and decision-making processes, not only to improve the accuracy of the experimental trajectory but also to increase trust by allowing transparent human-machine collaboration. High-quality structural characterization techniques (e.g., X-ray, neutron, or static light scattering) are a particularly relevant example of this need: they provide invaluable information but are challenging to analyze without expert oversight. Here, we introduce AutoSAS, a novel framework for human-aside-the-loop automated data classification. AutoSAS leverages human-defined candidate models, high-throughput combinatorial fitting, and information-theoretic model selection to generate both classification results and quantitative structural descriptors. We implement AutoSAS in an open-source package designed for use with the Autonomous Formulation Laboratory (AFL) for X-ray and neutron scattering-based optimization of multicomponent liquid formulations. In a first application, we leveraged a set of expert defined candidate models to classify, refine the structure, and track transformations in a model injectable drug carrier system. We evaluated four model selection methods and benchmarked them against an optimized machine learning classifier and the best approach was one that balanced quality of the fit and complexity of the model. AutoSAS not only corroborated the critical micelle concentration boundary identified in previous experiments but also discovered a second structural transition boundary not identified by the previous methods. These results demonstrate the potential of AutoSAS to enhance autonomous experimental workflows by providing robust, interpretable model selection, paving the way for more reliable and insightful structural characterization in complex formulations.

**Introduction**

The scientific and engineering community is in the throes of an artificial intelligence (AI) and machine learning (ML) paradigm shift. This is, in part, driven by the advent of automated high-throughput synthesis and characterization tools aimed at accelerating the optimization and discovery of new materials. These platforms, often referred to as material acceleration platforms[1] (MAPS) or self-driving labs[2] (SDLs), show great promise, with notable examples in battery[3], mechanical design[4], and other areas. Despite this, challenges remain for their large-scale adoption. Research groups developing and applying SDLs are seeking undeniably transformative breakthroughs in materials discovery while also trying to build trust in these methods and their results.[5,6] A key step in this trust building process is the development of interpretable, physics-based analogs to the black-box methods employed in other fields of AI and ML[7,8].

Autonomous experiments, MAPS, and SDLs all require rapid and reliable scientific analysis of measurement data and most efforts have focused on data driven strategies that use black-box machine learning[9] These strategies have been leveraged to varying degrees of success for classification tasks associated with spectroscopy, physical properties, scattering and other techniques.[10–13] Such methods provide fast and highly accurate classification when there are enough distinguishing features in the data or when the training set contains all possible presentations of the data expected in the experiment, but can fail to extrapolate or identify when presented with new or non-ideal data. Additionally, black-box models and techniques like neural nets are commonly employed for their speed and accuracy in the large-data limit, but can be difficult to train with the small amount of measurement data available for most materials. Finally, it is challenging to program black-box models with physical and chemical knowledge. This ultimately limits both the sophistication of the autonomous experiment and the possibility of human-machine teaming.

Next-generation autonomous experiments should be able to flexibly answer sophisticated scientific questions or conduct targeted optimization campaigns that depend on the physics of the system *e.g.,* map the bounds of a particular phase, identify all the unique structures within a material, or determine the compositional region where a structural property is in a target range. Structural characterization techniques are at the core of many scientific and engineering questions and an area where ML analysis has demonstrated difficulty[14] Small angle scattering (SAS) experiments present a particular challenge for application of ML methods for a variety of reasons: (a) a material's structural response to an external stimulus or composition change may vary continuously, (b) structural models for mostly disordered materials are more ambiguous than those typical of crystalline materials, (c) multiple material structures can produce identical or nearly-identical scattering patterns, (d) experimental and scattering artifacts can cause false structural features, etc. In the biological SAS community, quantitative model selection routines are increasingly incorporated into user-friendly packages such as BioXTAS RAW and ATSAS, but comparable turnkey solutions are more rare for materials SAS (though SASfit has some capacity). This gap likely arises from the fact that materials systems span a far wider variety of morphologies, size distributions, and hierarchical architectures than largely monodisperse biomolecules, so there exists a large pool of candidate models with highly correlated/entangled parameters. In view of these challenges, there is a clear need for automated data modeling techniques that leverage the last 60 years of expertise in the physics of SAS and incorporate that human expertise into autonomous experiments.

Here we present a new software tool, AutoSAS, that incorporates expert understanding in an autonomous experimental loop by starting from constrained, physically reasonable candidate structures and applying intelligent model selection to address the difficulty of fitting SAS data. The core of this fitting routine is built on top of two widely used software packages developed for SAS: *sasmodels*, a library of analytical SAS models, and *bumps*, a flexible optimization and fitting engine. The primary advantage of AutoSAS is that the automatically selected best model acts as a physically descriptive label, unlike the uninterpretable labels received from many black-box ML models. This approach further provides the ability to weigh the accuracy of model selection and the reasonableness of the pre-selected pool of models by using a "none of the above" category label for datasets which do not pass a goodness of fit threshold. We expect this tool to enable a new generation of SDLs that employ SAS as the driving analytical technique, and we expect our human-aside-the-loop strategy to form the basis of similar advances in other complex analytical techniques.

## Methods

### Solution Formulation

Samples containing solute mixtures of the surfactant poloxamer P188 (also called Pluronic F68), benzyl alcohol, and phenol components in deuterated and non-deuterated 5 mM pH 6 sodium phosphate buffer solvents were prepared manually for data collection. Stock solutions of approximately 30 mg/mL benzyl alcohol, 35 mg/mL phenol, and 100 mg/mL of poloxamer P188 were prepared following previous procedures[15]. The data collected and analyzed consist of semiregular sampling over the 3 solute components. All reagents were purchased from Sigma-Aldrich and used without further purification.

### Small-Angle X-Ray Scattering (SAXS)

The solute structures relevant to this surfactant-preservative system are on the order of 10 Å to 100 Å length scale, thus demanding small angle scattering as an authoritative measure of bulk structure.

SAXS data used in this study were collected at the BioSAXS facility at beamline ID7A1 of the Cornell High Energy Synchrotron Source (CHESS). This fully in-vacuum SAXS beamline is fed by a Cornell Compact Undulator and a multilayer monochromator, operated at a photon energy of 9.8 keV with typical flux of $7 \times 10^{12}$ photons in a 250 um x 250 um aperture. Samples were loaded in a custom flow cell and data collected on an in-vacuum Eiger 4M pixel-array detector operated at a typical sample-detector distance of 1.648 m. The sample-detector distance and beam center were calibrated using silver behenate and the intensity calibrated using water as a primary reference with cross-calibration performed to glassy carbon. 60 uL of sample was oscillated at a rate of 1-2 seconds in the instrument during each exposure to limit beam damage. Data were reduced and averaged using the BioXTAS RAW software[16].

**AutoSAS**

AutoSAS was developed in Python for the AFL platform[17] using a client-server interaction architecture is implemented in production in the AFL-agent repository[18]. Further supplementary notebooks and data are hosted in a dedicated repository[19]. AutoSAS is specifically implemented in a microservice, meaning a modular, small-scope, often stateless service that can be interfaced easily from other domains[20]. The benefits of encapsulation of components of autonomous experimental platforms in microservice architectures are wide-ranging, from separation of duties to the ability to reconfigure and restart individual servers without restarting an overall, long-running loop. For fitting specifically, one crucial advantage of the HTTP microservice approach is that compute-intensive tasks (like parallel fitting using expensive approaches like Markov-chain Monte Carlo) can be hosted on high-performance and/or cloud computing resources with easy access over local networks or even the Internet to the actual experimental hardware. The core fitting engine uses the sasmodels package developed by the SASView community with access to programmatic modeling and fitting capabilities. A fitting server is started with a configuration file containing model information defined by an expert using sasmodel nomenclature including the model type, initial parameters, bounds over which those parameters can be varied if applicable, and a corresponding scattering vector (q) range for that model. Additional settings include specification for the fit method with corresponding parameters (i.e., Levenberg-Marquardt (LM), number of iterations, tolerance threshold, etc.). Experimental data in the form of q, I(q), dI(q) and dq(q) are provided in a network call and causes the server to fit each established model. Here, q, I(q), dI(q) and dq(q) are the scattering wave vector, the scattering intensity, the standard deviation of the scattering intensity, and the width of the q resolution of an instrument, respectively. The results of each fit, including the converged parameters, the $\chi^2$ of the fit, estimate of model probability (if using the probabilistic method), and the best model selection are returned in a JSON format which combines machine and human readability: the core of AFL-automation can also store this data in a Tiled server with rich metadata tagging, allowing retrospective parameter tracking and fit reconstruction. The structure of this fitting architecture and its role in an autonomous decision-making loop is shown in Figure 1.

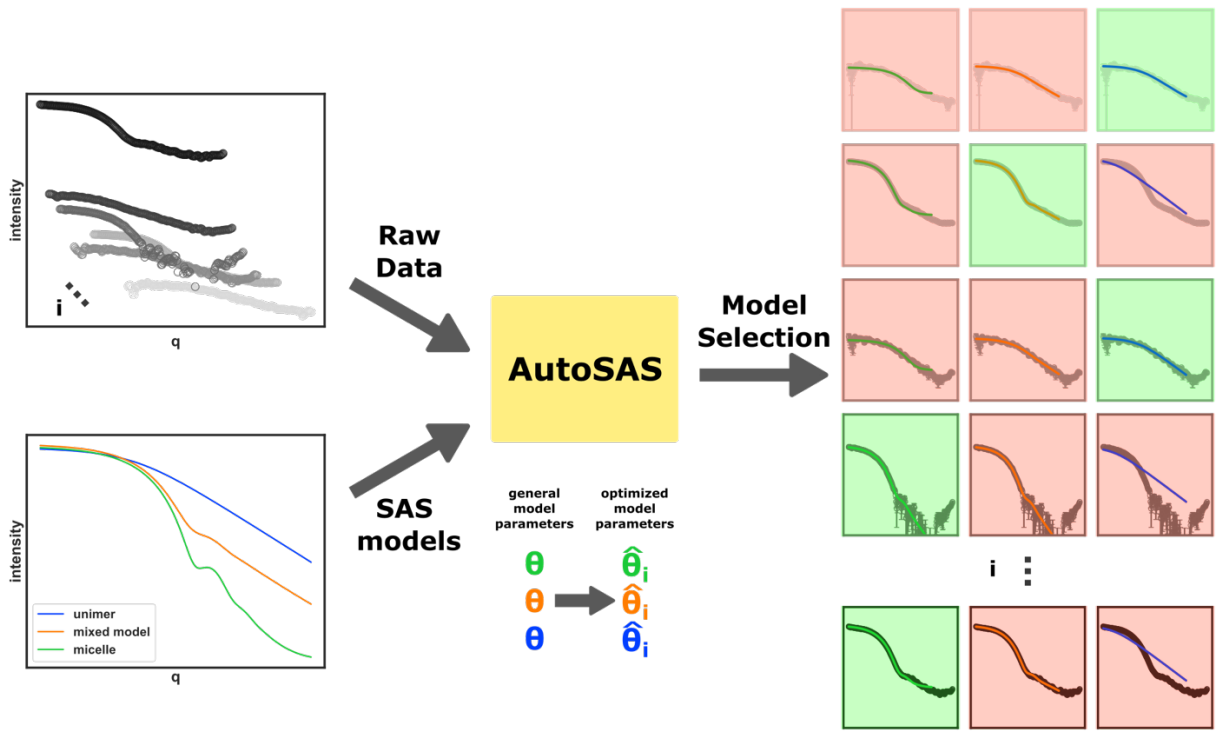

**Figure 1.** Schematic of the AutoSAS architecture where data and constrained SAS models are passed as inputs to the AutoSAS module. All combinations of data and models are optimized, followed by model selection under some pre-defined criteria. The output is the complete set of optimized models with a best candidate model (indicated by the green box) and rejected models (red box).

**SAS models**

A wide variety of SAS models have been developed and here we will briefly describe those relevant to our demonstration. Some analytical form factor models describe the scattering particle shapes from first principles (*e.g.,* sphere, cylinder, ellipsoid, core-shell variants, etc.) but come with many fitting degrees of freedom that can over-represent the data. Such models are often a poor representation for non-colloidal materials owing to the multiscale nature of their structure, and for polydisperse or heterogeneous mixtures. More phenomenological models can contain fewer (yet still physically meaningful) parameters, which can be coupled with prior knowledge and other techniques to characterize complex and/or non-equilibrium, evolving structures. Though any arbitrary set of models could be fit using the AutoSAS approach, we limit our consideration for this demonstration to four models, driven by expert understanding. "Ground truth" datasets in this study are the result of comprehensive fitting using the simple Guinier's law[21], a robust and simple method for extracting radius of gyration and size of the scattering particles.

*Guinier Model*

$$I_{Guinier}(q) = I_o \cdot \exp\left(\frac{-q^2 R_g^2}{3}\right) + background$$

The three parameters in this model describe the radius of gyration $R_g$ of a particle and the pre-factor $I_o$ indicates the mass and concentration. This model works at low q scattering vectors but relies centrally on an assumption of monodisperse, non-interacting particles, i.e., particles at low concentrations. Aggregation and particle-particle interactions can cause this to deviate.

*Spherical Form Factor*

$$I_{sphere}(q) = \frac{scale}{V} \cdot \left[3V(\Delta\rho) \cdot \frac{\sin(qr) - qr\cos(qr)}{(qr)^3}\right]^2 + background$$

This geometrically derived form factor model[21] produces a scattering pattern assuming a spherical geometry with a uniform scattering length density (SLD) distribution, where $\Delta\rho$ is the difference in SLD between the particle and solvent, r is the radius of the sphere, V is the volume of the sphere in solution, and *scale* is a weighted volume fraction if SAS data are not reported on an absolute scale. Some degree of polydispersity can be treated by summing across a fixed number of bins of allowed parameter, e.g., radius.

*Polydisperse Gaussian Coil Model*

$$I_{poly\_gauss\_coil}(q) = scale \cdot I_0 \cdot P(q) + background$$

A polydisperse Gaussian coil is another form factor model[22] that describes a polymer-like structure by assuming that segments of a linear polymer chain follow a random walk. The form factor term P(q) also includes the polydispersity (assuming a Schulz-Zimm distribution of molecular weights), the volume fraction of the scattering particle, and the radius of gyration.

*Mixed model*

$$I_{mixed\_model}(q) = \left(scale_A \cdot I_{poly\_gauss\_coil}(q) + scale_B \cdot I_{sphere}(q)\right) + background$$

The "mixed model" used in this work is a scaled composite of the polydisperse Gaussian model and the sphere. There is a single background term and two scale terms that fractionalize the contributions from different type of particles in solutions.

**Model Selection**

In our use-case, the overarching challenge of model selection is a tradeoff between the level of physical detail in each model and the structural information contained in each set of SAS data. In an automated decision-making process, a critical step for SDLs or autonomous experiments, this becomes particularly important. The AutoSAS approach is to start from a set of expert chosen models with constrained parameters, automatically fit all models to all measurements in a dataset, and, by some selection method, choose the most appropriate model representing the data. This pre-definition of candidate models helps address some of the issues of fitting SAS data, specifically the many degenerate representations of data that can be produced from different models.

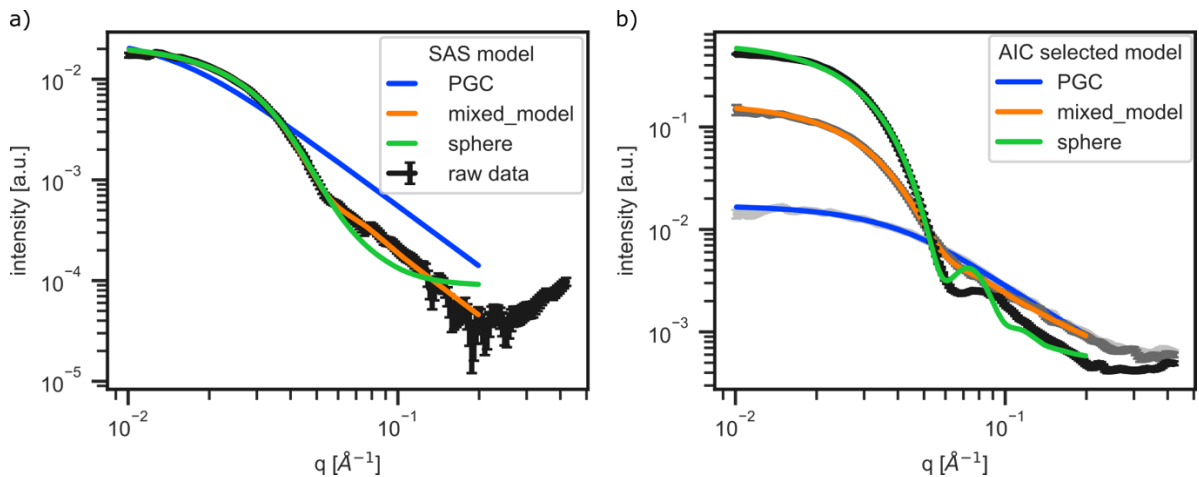

**Figure 2.** a) Representative SAXS data from the P188-preservative system with the best fit candidate structures after LM optimization. b) The evolution of the observed SAXS data and corresponding form factor SAS model selected using the Akaike information criterion (AIC) with increasing phenol concentration for a fixed amount of P188 and benzyl alcohol.

A representative SAXS pattern and the three competing structural models, defined above, with corresponding best fit results are shown in Figure 2a. Here, all models were optimized via the LM algorithm, where the mixed model (orange) best represents the data and shows the lowest $\chi^2$ value, our goodness of fit (GOF) metric. The sphere model (green) is also a reasonable fit for much of the q-range of the data and should be considered in the decision-making process. For all models, the fit quality at higher q-values should be considered less strongly as the data has lower intensity, lower signal to noise and is more sensitive to the background subtraction process.

The ability to rapidly and reliably fit is not enough for an autonomously driven experiment. Choosing a model from the user supplied set of models, or when to flag "all models are bad", is a challenge that balances model descriptiveness and data quality, *i.e.,* over/under fitting. Which model most appropriately represents the data must be determined automatically by some selection method which can utilize any combination of the goodness of fit, optimal parameter values, parameter uncertainties, or human bias and intuition. We demonstrated the importance of automated model selection from an information theoretic approach, an effective encoding of an AI's "thought process", by comparing the following four increasingly complex decision-making methods.

The first criterion was simply to choose the model with the lowest $\chi^2$ value, mathematically defined in Supplementary Information Section S1.1. This unitless metric is simply the average mean-squared difference between the model and the data normalized by the variance, where a lower value implies a better fit. If the $\chi^2$ value is 1, on average the data fit the model to within one standard deviation of the data's experimental uncertainty. The advantage of using this criterion is the straightforward interpretation: the model that most closely reproduces the data is the most appropriate. This is a reasonable approach when contending models are strongly dissimilar from one another.

The second criterion was a modified Occam's razor approach, which balances $\chi^2$ with the number of varied parameters, giving preference to simpler models in cases where the $\chi^2$ was comparable. A model with a slightly higher $\chi^2$, but with fewer varied parameters, is selected when the difference between the $\chi^2$ values between models is less than some user supplied value, here that threshold is 1. This heuristic is an arbitrary limit on the tradeoff for GOF to overly expressive models but produces a sensible result with an interpretable statistical limit.

The third selection method was to choose the model with the lowest Akaike information criteria (AIC). AIC is a statistical metric that is comprised of the GOF and a penalizing term for the complexity of the model under consideration (see SI Section S1.2), *i.e.*, the lowest $\chi^2$ value and number of varied parameters used in describing that model.[11] At its core is the information tradeoff between how much better one model can represent the data over another given the increase in complexity.

Our fourth and final method of model selection was a probabilistic approach used in analyzing X-ray diffraction data for crystalline materials[24] In this method, we estimate a probability for how well a model fit the data using the GOF, uncertainty estimates for the fit parameters, and the total number of parameters varied in every model proposed. The mathematical details for the probability estimates are reported in the SI Section S1.3. This method, like the others, balances the GOF and complexity in the number of varied parameters, but it also accounts for the certainty in optimized parameters. A model is additionally penalized if varied parameters are uncertain, not contributing to the quality of the fit.

The values for the AIC and probabilistic methods alone do not imply the quality of the model to fit the data but is discriminating when comparing different models. If the GOF is bad for all models, these metrics do not help you select a correct one, rather, this implies that an appropriate model is not in the user supplied set. As noted above, for this dataset at least one of the models was appropriate for the experimental data, so we use the direct values as the selection criteria.

Figure 2b shows the results for 3 different patterns in the dataset using the AIC selection method. As the scattering pattern changes with solution composition, AutoSAS picks the most appropriate model giving a physically informed label and the corresponding fit parameters. This kind of information is used for classification and future regression tasks that underpin autonomous experiments. The descriptiveness from both the model identity and parameters allows a scientist to engage with the AI in a more scientifically meaningful way.

**Results and Discussion**

To test the AutoSAS paradigm on a real-world dataset, we fit experimental SAXS data from a previously investigated 3-component surfactant-preservative pharmaceutical bioformulation.[15] The formulation consists of a poloxamer block copolymer, P188, with two common preservatives, phenol and benzyl alcohol, a common "framework" formulation in intramuscular protein therapeutic injections (e.g., insulin)[25] The system's mesoscale structure undergoes several morphological rearrangements as preservative and surfactant concentration are varied, chiefly between free surfactant in solution, a surfactant micelle with preservative cargo, and a turbid aggregate. The identity of the structures or phases and correspondingly, the transformations, are important to track in reformulation design. Recently, it was observed that the interplay between preservatives and surfactant molecules cause a structural transition of the molecules, transforming from the unimer to micelle phase. The surfactant concentration needed for this transformation is called the critical micelle concentration (CMC). These transformations may influence the efficacy of the excipients and ultimately the shelf life of the pharmaceutical. However, the complexity of the phase behavior due to multi-component compositions represents an important and real challenge for efficiently quantifying and mapping the phase boundaries in sophisticated autonomous experimentation. Beyond accurately fitting the SAXS data, AutoSAS needs to automatically decide which of the expert supplied candidate models is most appropriate.

We begin our analysis by assessing how well AutoSAS reproduces the results from the SAS expert's interpretation. One reasoned, "expert" approach to interpreting the SAXS dataset is *via* a Guinier analysis which can estimate the size of solutes in a solution by analyzing the behavior of scattering data at the lowest measured q-values. After fitting the Guinier model to the data, the corresponding ground truth phase labels can then be determined by the uptick in the parameter $I_0$ from the Guinier model, see Figure S2.2. Solutions with scattering below the average $I_0$ at low concentrations of phenol correspond to a unimer region, and above that indicates micellization. With this definition and analysis, we can label the data and define this as the expert 'ground truth' labels. We use these labels to assess the performance of AutoSAS.

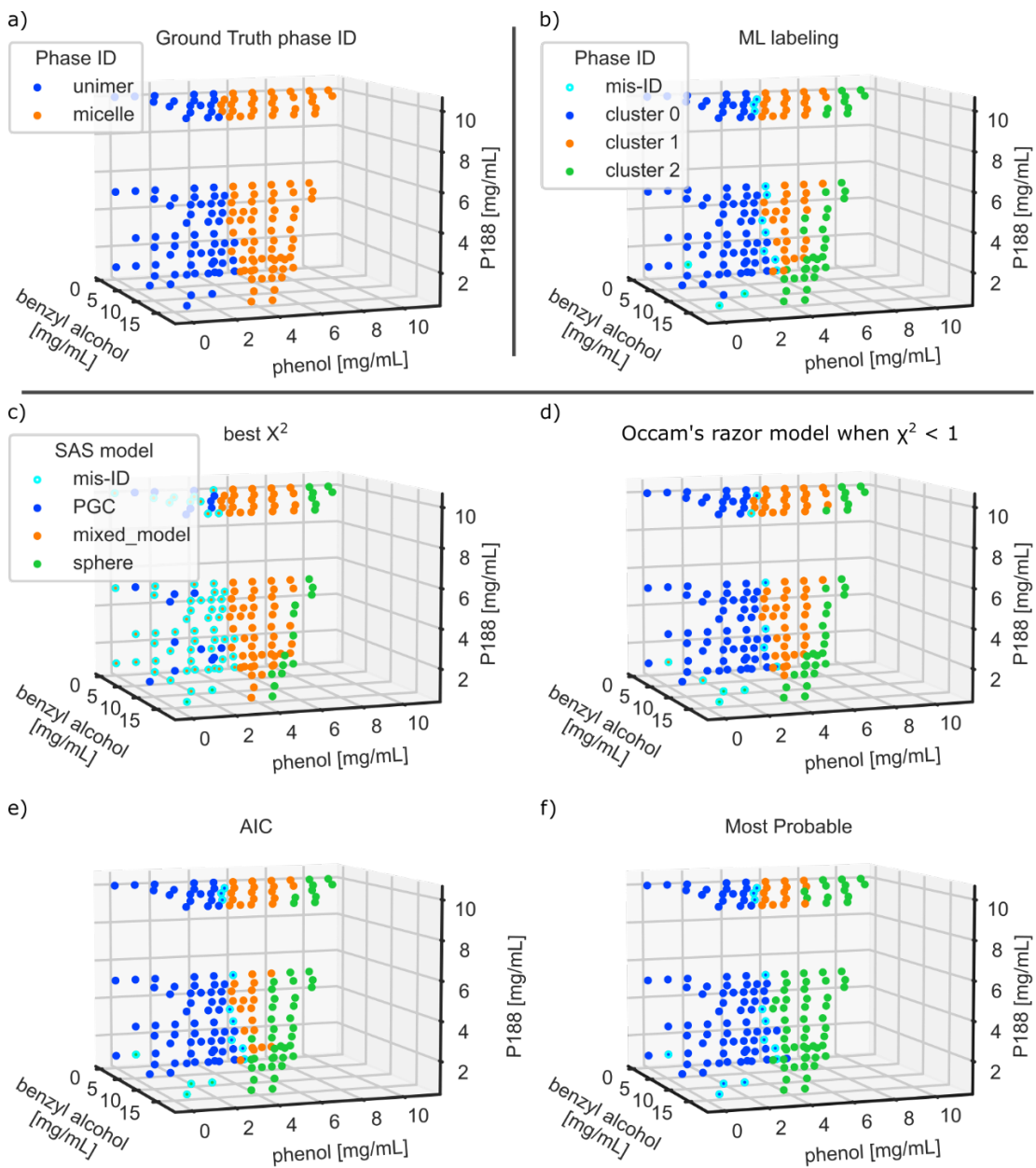

**Figure 3.** Classification labels from ~~the~~ a) the ground truth phase identification, b) the ML classification method, c) the lowest $\chi^2$ approach, d) Occam's razor, e) highest AIC, and f) ~~a~~ probabilistic method.

Figure 3 compares the four AutoSAS labeling methods described in the "Model Selection" section above against the expert ground truth labels and the labels from a pure data approach, labeled as ML. For the ML approach, we use a previously developed SAS classification scheme described in full detail in ref. 26. This method first calculates the Laplacian similarity scores for the 0[th], 1[st], and 2[nd] order derivatives of the scattering data and then uses a spectral clustering algorithm to group the data into classes. The hyperparameters were determined using a large amount of simulated SAS data with real instrument resolution functions and noise injection, see reference. We chose this method as a benchmark based on its performance in our past ML-driven SAS analyses. Of course, there could be more performant clustering or classification algorithms, so this ML approach is better considered an indicative benchmark for an ML-focused strategy rather than a definitive "best". This method works well when there are discrete phase boundaries with large differences in the scattering patterns, but can 'hallucinate' phase fields in a continuously changing boundary (e.g., the onset of micellization from unimeric polymers, where the intensity of the strong micelle scattering varies continuously across the boundary). This method also requires prior knowledge about the number of phases/clusters or a secondary analysis to estimate this number. We employed a "best-case scenario" for this algorithm by prespecifying a maximum of 3 phases. The results of the ML classification are shown in Figure 3b where each color marker corresponds to a different category.

For the AutoSAS fitting and labeling, we applied relatively simple form factor models that are motivated by the physics of the material system under study. See the examples of the fit ranges and quality in Figures 2a and 2b. The first model (blue curve) was the polydisperse Gaussian coil (PGC), representing the free unimeric surfactant in solution, with parameters of $R_g$, scale, and background. All other parameters in this model were fixed to known values and are reported in the fit value tables in Figure S6.1. The second AutoSAS model was a spherical form factor, representing the micelle (green curve), with a radius, radius polydispersity, scale, and background, again with fixed and known parameters. We introduce a third, mixed model (orange curve), defined by as the sum of both the polydisperse Gaussian coil and sphere form factor models. This mixed model contains similar parameters described above, with the exception being a single background term and two separate scale terms for each component to allow for this continuous transformation. This mixed model represents coexistence of the unimeric and micelle-like structures in solution, consistent with the physics of the critical micelle concentration transition. These models provide a physical and geometric understanding of the evolving solute structure as a function of the solution composition.

AutoSAS determines the classification label by fitting each of the three models described above and then applying one of the four selection criteria detailed in the Methods section. The results of this classification are shown in Figure 3b-f and in more detail in Figures S3.1-3.6. In the AutoSAS analysis, the color of the marker indicates the model chosen by the selection method while the presence of a cyan border indicates disagreement with the ground truth. We observe that the $\chi^2$ approach (Figure 3c) performs the worst of the AutoSAS methods, with many misclassifications in the unimer region. Comparatively, the other three AutoSAS approaches seem to perform similarly, with the majority of misclassifications occurring near the unimer to micelle transition boundary.

**Table 1.** Fowlkes-Mallows Score between model selection and ground truth

| # of clusters | lowest $\chi^2$ | Occam's razor | AIC | most probable | ML |
|---|---|---|---|---|---|
| 2 cluster | 0.717 | 0.780 | 0.780 | 0.800 | 0.809 |
| 3 cluster (merged, see text) | 0.738 | 0.877 | 0.856 | 0.836 | 0.856 |

We can quantitatively compare the results of the classification to the expert labeled ground truth by the Fowlkes-Mallows score (FMS), where a value of 0 indicates all data are incorrectly labeled and 1 indicates perfect assignment (Table 1). The first row gives the FMS as a function of selection method with AutoSAS fitting only the unimer and micelle models and the ML model fixed at two clusters. This selection condition is used for direct comparison between the Guinier analysis ground truth (which contains only two clusters) and the ML and AutoSAS methods. While Guinier analysis provides an unambiguous and independent mathematical ground truth, it misses subtle known physics of the materials system, e.g. the coexistence of unimers and micelles above the CMC. The best available analogous three phase "ground truth" would be an analytical fit identical to that performed by AutoSAS itself; in light of this, to benchmark the performance we compared the full AutoSAS candidate set of three models and ML fixed to 3 clusters, and considered only the boundary between the merged (mixed + micelle) and unimer models. Comparing these combined-phase cases to the Guinier two-phase ground truth allows us to benchmark a more physically realistic version of the two models. We caution that this approach suffers from some label-cardinality bias, but we believe that it is the best realistically achievable comparison in the absence of explicit three-phase ground truth.

As discussed above with a visual analysis of Figure 3, the lowest $\chi^2$ method produces the worst results (Figure 3c) as it tends to prefer the mixed model in the unimer region at low concentrations of P188 and phenol. This poor representation is likely due to noisy scattering patterns appearing to fit better with the most flexible model, a classic case of overfitting. Because this approach overfits the mixed model, we do not consider the $\chi^2$ approach for the remainder of this work. The three more sophisticated selection methods produce better classification with the modified Occam's razor approach performing the best, followed by AIC, and the most probable model comparison. The ML classifier performs as well as the AIC method, albeit with carefully selected hyperparameters and knowledge about the number of clusters to separate the data.

The more sophisticated methods tend to misidentify a few scattering patterns at the border between the unimer to micelle transition and some in the low concentration P188 and phenol regions. In both the AIC and Occam's razor methods, about half of the incorrectly identified phases are at low concentrations of P188 and phenol where the patterns have low signal to noise levels. The probabilistic method handles the noisier data better but struggles at the CMC boundary.

The discrepancy between classification methods and the ground truth is likely due to the sensitivity in defining the CMC boundary. The structural transition based on the Guinier analysis is biased towards where there is a significant initial change in the $I_o$ parameter, whereas the ML method is affected by where the difference between scattering patterns is largest. If the $I_o$ intensity is not changing, but the shape of the curve is, then the Guinier analysis will disagree with a model specific classification. AutoSAS is affected by a similar sensitivity to the shape of the SAS pattern and the tolerance of the selection criteria. These locations in composition space are not guaranteed to coincide, and indeed, a different scientific problem might drive a desire to measure the micelle onset using either an $I_o$ based approach or a more statistically weighted scheme based on fractional conversion of unimeric surfactant to micelle.

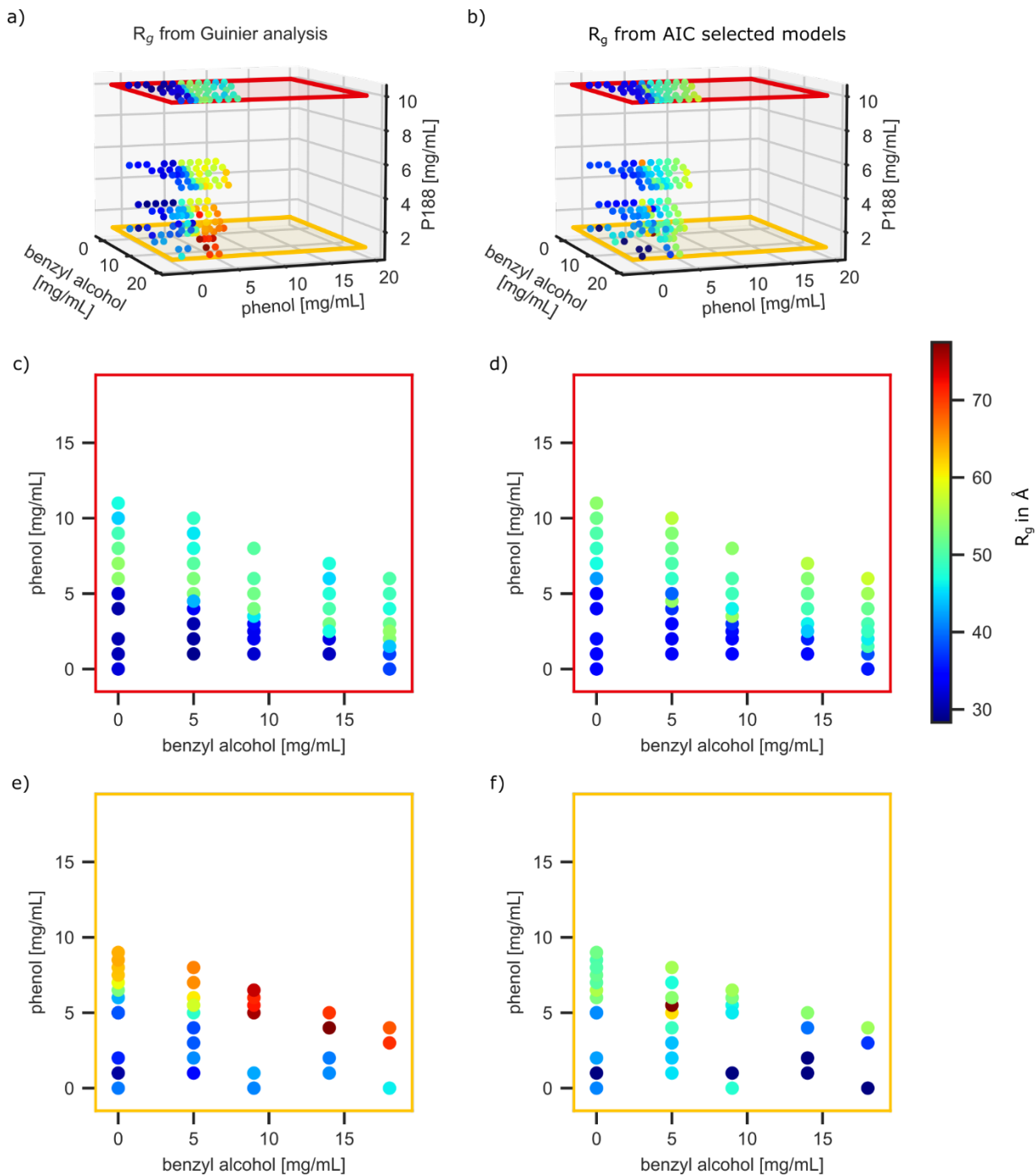

**Figure 4.** a,c,e) Guinier analysis ground truth results of the $R_g$ parameter as a function of the surfactant-preservative compositions. b,d,f) the AutoSAS fitting results for AIC selected models.

The critical advantage of the AutoSAS approach is that it provides additional structural information that the ML classifier cannot. In this system, for instance, the comparative effects of addition of phenol versus benzyl alcohol into the micelle can cause subtle changes in the micelle size. To demonstrate this, we compare the fit values of $R_g$ and effective $R_g$ in the models selected by AutoSAS to the values from the ground truth. Plots of effective $R_g$ over all solution compositions for the Guinier and composite models are shown in Figures 4a and 4b, respectively. Figures 4c-f show 2D slices at the highest and lowest concentrations of P188 for easier comparison with all other compositions in Figure S2.5. There is a sharp increase in $R_g$ while moving across the CMC boundary in both analyses. This behavior is generally captured by AutoSAS, although there are more outliers at lower concentrations of P188 and the size of the unimers are identified to be larger than that of the Guinier analysis. There is an anomalous increase in $R_g$ at lower concentrations of P188 in the micelle region from the Guinier analysis. This may originate from low signal to noise in the scattering data, particularly at very low and high q values, and background contributions, which are known limitations of this analysis.

The use of the mixed unimer-micelle model (*vide supra*) elucidates a *second, practical boundary* that cannot be readily identified by the Guinier analysis, as there is no discontinuous change in the $I_0$ parameter. This boundary is the point at which micelles become the volumetrically dominant scattering object even though unimers are still present. For clarity we use the shorthand "micelle-conversion threshold" (MCT) but must stress that it is not a new thermodynamic transition; it is simply a convenient, operational marker derived from our model-based estimate of micelle fraction. Crossing the MCT has design significance because the residual unimer population, and therefore any unimer-specific functionality, falls below levels detectable by scattering. To illustrate how the data evolve across this boundary, we plot in Figure 5 the mean (solid line) and variance (shaded region) of the ML-classified clusters. Three characteristic signatures appear: a low-intensity, $q^{-2}$ regime for predominantly unimeric solutions (blue); a $q^{-4}$ regime for spherical micelles (green); and an intermediate form (orange) that bridges the two. The discriminating features are the overall intensity and the mid-$q$ slope change (0.03–0.11 Å$^{-1}$). These observations justify our earlier choice to merge the intermediate and micellar clusters when estimating micelle fraction, while the shaded bounds highlight intensity variations that track most strongly with P188 concentration.

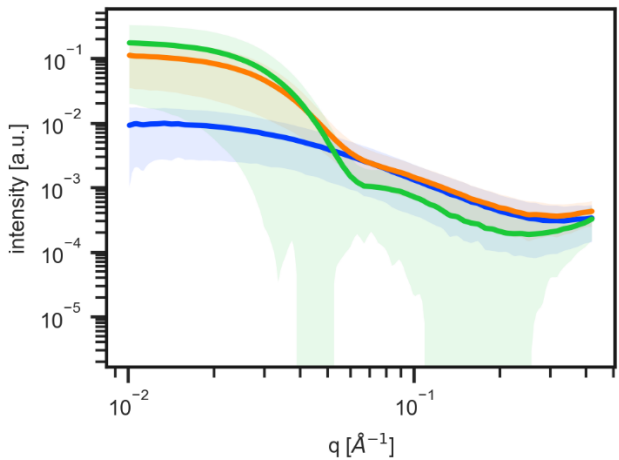

**Figure 5**. The mean and uncertainties of the SAXS data clusters as identified by the ML classifier

An additional advantage for using the mixed model is determination of what the relative fractions of micelles and unimers are in solution, also providing a quantitative evaluation of our hypothesis on the nature of the MCT boundary. We calculate the relative fraction of micelles to total polymer solute in solution within the mixed model domain, just after the CMC threshold, and just before the MCT. We construct a volume fraction of micelle solute using the scale parameters from the constituent models in the mixed model selected regions. This is defined as the scale of the sphere model divided by the sum of the sphere scale and polydisperse Gaussian coil scale terms. Figure 6a-d shows the evolving fraction of micelles in solution over all compositions with the CMC and MCT boundaries indicated by the cool and warm colored interpolated surfaces by the more sophisticated selection methods. 2D projections at fixed P188 compositions with the boundaries are plotted in Figure S4.1 – S4.5.

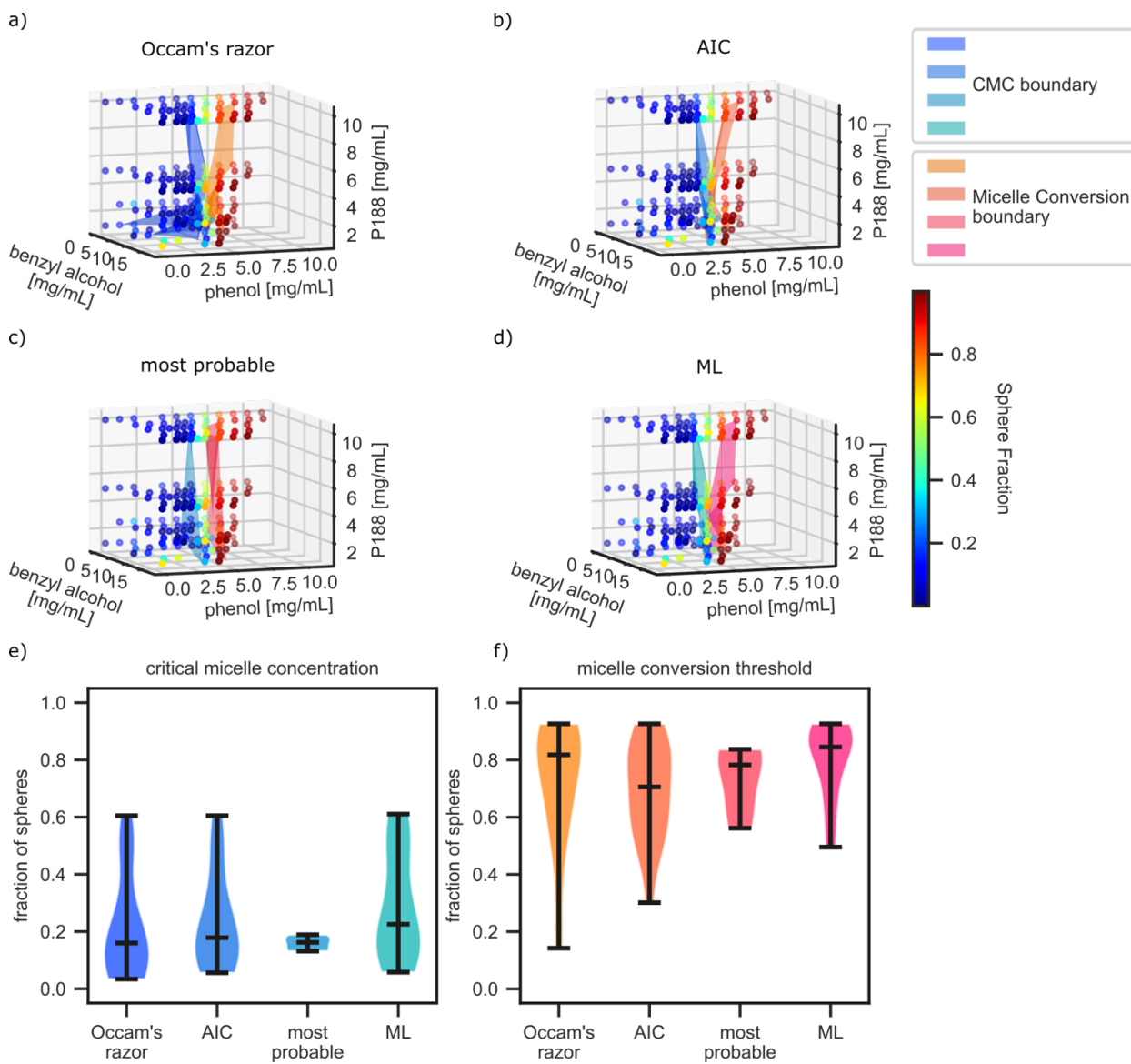

**Figure 6.** The fractional distribution of spheres in the mixed model over all compositions with the CMC and micelle conversion boundaries indicated by the splined surfaces for a) Occam's razor, b) AIC, c) most probable, and d) ML selection criteria. e,f) Violin plots indicating the fraction of micelle just above the CMC and just below MCT transitions as determined by the mixed model.

Figure 6e and f show violin plots for micelle fractions just above the CMC boundary and just below the MCT boundary determined by the different selection criteria and the ML classifier. The violin plots are histogram representations of the micelle fractions with the inter quartile regions and medians marked by the black bars with a smoothly interpolated histogram between. The data in these plots originate from the boundaries of the mixed model identified region. There is a spread in the distributions, which we attribute to the erroneously classified patterns, but the median values corroborate an upper limit to the beginning, and lower limit to the end of, the unimer to micelle transformation, summarized in Table 2.

Table 2: Fraction of solute micellized at midpoint of the structural transition

| Boundary | Occam's razor | AIC | most probable | ML |
|---|---|---|---|---|
| CMC | 0.16 | 0.22 | 0.30 | 0.29 |
| MCT | 0.82 | 0.70 | 0.95 | 0.88 |

The Occam's razor selection method identified the largest difference between the CMC and MCT with the narrowest distributions around each. Applying a similar analysis to the AIC selection method produces broader distributions of micelle fractions with the transformation boundaries together. The probabilistic method has the broadest distribution for the CMC and MCT boundaries but also had fewer data points to define these transformations. The median values are consistent with the lowest AIC and modified Occam's razor methods. Lastly, the ML method produces a distribution of fractions like both the Occam's razor and AIC methods for the CMC boundary but has a tighter distribution of concentrations for the MCT with median values in line with the other methods. The span of what is identified as the "mixed model" can be seen clearly in Figure 5, providing insight into what is considered the CMC or the MCT. The Occam's razor, lowest AIC, and ML methods produce results that have more contiguous domains which reduce uncertainty and span of the threshold distributions compared to that from the probabilistic method. One reason for the inconsistency with the probabilistic method is that the penalizing terms from the Laplace approximation are too severe, causing comparatively high probabilities for the pure sphere model as opposed to the mixed model. Stated another way, there is not enough evidence for the mixed model given the uncertainties in the data from the fit values, driving the choice for the simpler model.

It is interesting to consider how the AutoSAS approach would scale to a larger number of models and to models with increasing complexity. Because model fitting is intrinsically parallelizable across datasets and models, computing power alone is unlikely to be a serious constraint. A greater challenge arises in deciding which model is appropriate when many models reasonably describe the data, a common occurrence in scattering. One approach would be the inclusion of information from complementary techniques (multi-angle light scattering, UV-vis spectroscopy, size exclusion chromatography, etc.) to help distinguish similar models. Ultimately, though, the central premise of the AutoSAS paradigm is that one need not fit *all possible models in the universe*; rather the human aside the loop predefines a *reasonable* number of physically informed candidate models for the material at hand. In some sense, the computational and probabilistic limitations described above will rarely be a limitation to a real study.

**<u>Conclusion</u>**

We envision AutoSAS as the starting point for interpretable and flexible autonomous experimentation in structure driven studies using small angle scattering. Furthermore, AutoSAS serves as a template for deploying quantitative model selection strategies for other measurements in studies from across the scientific spectrum.
AutoSAS was built around accepted community analysis software, implemented in Python, and designed for state of the art automated and autonomous experimental use. We leverage fitting engines applied to a set of expert defined models capable of incorporating instrument artifacts and resolution functions from real experiments into the analysis. The results of the fitting allow us to automatically build a scientifically accurate description of the structural space, given a hypothesis of what set of structures is likely to exist. Moreover, AutoSAS employs intelligent automated model selection to choose the candidate that best represents the data, overcoming some of the challenges of fitting SAS data. Making this decision based on a hypothesis that can be tested is difficult to implement with purely ML based approaches. Having quantitative metrics for GOF and selection criteria diversity allows us to flag when models or a hypothesis are incorrect.

We quantitatively benchmarked the ability of AutoSAS to reproduce an expert derived ground truth on experimentally collected SAXS data for the poloxamer P188 surfactant-preservative system. We evaluated four different selection criteria, the model with lowest $\chi^2$, a modified Occam's razor approach, Akaike information criteria, and a probabilistic method from AutoSAS and an optimized machine learning approach to classify the SAS data. These were quantitatively compared to the ground truth for the critical micelle concentration boundary identified using the Guinier method. We found that a modified Occam's razor model selection method was the most appropriate for hypothesis testing and accurately captured the ground truth classification. The AIC and probabilistic model comparison methods performed slightly worse for both tasks, although they might perform better for a different model set or classification task.

Beyond validation, we used AutoSAS to identify a second practical structural transition in the P188 material system which we denote as the micelle conversion threshold. This boundary is related to the end point in unimer to micelle conversion, which is not identifiable with the simpler

Guinier analysis used in previous studies. Increasingly complex fitting alone was insufficient to make this claim, but relied on the validated model selection methods implemented in AutoSAS to overcome the issues of overfitting.

In summary, AutoSAS is a physics-based modeling and learning approach that allows researchers to engage with an automated decision-making process in a way that is rational. Leveraging human insight and prior knowledge with powerful ML methods is the next and best way forward in AI accelerated science.

## Supplementary Material

Detailed mathematical derivations of the model selection criteria (chi-squared, AIC, and probabilistic methods), comprehensive fitting results and parameter comparisons between AutoSAS and expert Guinier analysis, phase identification results for all selection methods, visualization of CMC and MCT boundary transformations across all compositions, and complete fitting parameters for all three form factor models used in this study.

## Acknowledgements

The authors thank Richard Gillian for assistance with SAXS measurements. The authors thank the nSoft consortium at NIST for financial support. This work is based on research conducted at the Center for High-Energy X-ray Sciences (CHEXS), which is supported by the National Science Foundation (BIO, ENG and MPS Directorates) under award DMR-2342336, and the Macromolecular Diffraction at CHESS (MacCHESS) facility, which is supported by award 1-P30-GM124166-01A1 from the National Institute of General Medical Sciences and the National Institutes of Health and by New York State's Empire State Development Corporation (NYSTAR). Certain commercial equipment, instruments, materials, suppliers, or software are identified in this paper to foster understanding. Such identification does not imply recommendation or endorsement by the National Institute of Standards and Technology, nor does it imply that the materials or equipment identified are necessarily the best available for the purpose.

## Author Contributions

PAB, DRS, and TBM conceived the idea and designed the AutoSAS approach. RF prepared the samples and collected SAXS data. YL provided input into the model choices. DRS wrote the AutoSAS code, performed the analysis, and wrote the first draft of the manuscript. PAB and TBM assisted with manuscript refinement and supervised the work. All authors have read and approved the final version of the manuscript.

*Supplementary Information for*

# AutoSAS: a new human-aside-the-loop paradigm for automated SAS fitting for high throughput and autonomous experimentation

Duncan R. Sutherland[1,2], Rachel Ford[3], Yun Liu[3,4], Tyler B. Martin[1*], Peter A. Beaucage[3*]

[1]Materials Science and Engineering Division, National Institute of Standards and Technology, Gaithersburg, MD 20899, United States
[2]Department of Physics, University of Colorado at Boulder, Boulder, CO 80309, United States
[3]NIST Center for Neutron Research, National Institute of Standards and Technology, Gaithersburg, MD 20899, United States
[4]Department of Chemical Engineering, University of Delaware, Newark, DE 19716, United States
* to whom correspondence should be addressed, TBM: tyler.martin@nist.gov, PAB: peter.beaucage@nist.gov

Table of Contents

## S1. Model Selection Criteria

### S1.1 Best chi-squared

Which model, $m_i$, most appropriately represents the data is determined automatically by some selection method which can utilizes any combination of the goodness of fit (GOF), $X^2_{m_i}$, optimal parameter values, $\hat{\theta}_{m_i}$, parameter covariances, and $\Sigma_{\hat{\theta}_{m_i}}$. The GOF of each model is defined as the average square of the residuals between the model, $x_j \hat{\theta}_{m_i}$, and experimental data, $y_j$, normalized to the square of the experimental uncertainty, $\sigma_j$, over all scattering vectors $\{q_1, \ldots, q_n\}$, equation 1.

$$X^2_{m_i} = \frac{1}{n} \sum_{j=1}^{n} \frac{(y_j - x_j \hat{\theta}_m)^2}{\sigma_j^2}$$

(1)

### S1.2 Akaike Information Criterion (AIC)

AIC is defined as the log of the likelihood $\hat{L}_{m_i}$ plus the number of varied parameters, $k_{m_i}$, for a given model in equation 2.

$$AIC_{m_i} = 2k_{m_i} + 2\log(\hat{L}_{m_i})$$

(2)

The AIC is favored for its simplistic interpretation that penalizes complex models, like the Occam's razor approach, but derived in a more formal likelihood framework[1], originating from minimizing the Kullback-Leibler (KL) divergence between the optimized model and the data. Here the likelihood, $\hat{L}_{m_i}$, is the GOF value after AutoSAS fitting. The AIC value for a given model can be used for hypothesis testing and to discriminate models with other heuristic approaches if needed, but for this work we use the direct value as the selection method.

### S1.3 Estimating Probability

The estimating probability method requires knowledge of the model evidence, also known as the marginal likelihood which can be difficult or impossible to determine analytically. We use the Laplace approximation [2] for estimating this integral, a common approach in statistical analysis that does not require expensive Markov Chain Monte Carlo (MCMC) methods, as represented by equation 3.

$$\log\left(P_{m_i}(\boldsymbol{y}|m_i)\right) = \log\left(\int P_{m_i}(\boldsymbol{y}|\theta_i) P_{m_i}(\theta_i|m_i)\, d\theta_i\right)$$

$$\approx \frac{k_{m_i}}{2}\log(2\pi) + \frac{1}{2}\log|\Sigma_{m_i}| - \hat{L}_{m_i}$$

(3)

Equation 3 contains three terms, one for the complexity of the model, one to estimate the observed information which is a measure of parameter importance, and lastly the overall likelihood.

We calculate the probability of a model with the marginal likelihoods in equation 4.

$$P_{m_i} = \frac{P(\mathbf{y}|m_i)}{\sum_j P(\mathbf{y}|m_j)}$$

(4)

It should be noted that the Laplace approximation is only appropriate when there is a "peak" and unique optimum in the parameter space $\theta$ allowing for the estimate of the integral. If this is not met, which is hard to determine a-priori, the method for calculating the probability can fail. For a more robust and accurate determination of this likelihood, we can use the directed evolution / DREAM method implemented in the bumps fitting engine to estimate the parameter distributions at the cost of speed of standard Levenberg-Marquardt (LM) fitting. This study does not implement the DREAM method for fitting but the capability is built in.

## S2. $R_g$ comparison

We extract the $R_g$ and $I_0$ parameters, (Figure S2.1 and S2.2), and show that we can reproduce the ground truth (Figures S2.3 and S2.4) with high fidelity. There are a few outliers, observed in the $R_g$ parameter, which are affected by the range of data over which the model is applied. The Guinier model is appropriate in the limit as the scattering vector approaches small values. Decreasing the range of q-values too much for the fit increases the agreement of the $R_g$ with the ground truth but produces more outliers due to noise in the scattering patterns, particularly at lower concentrations of P188.

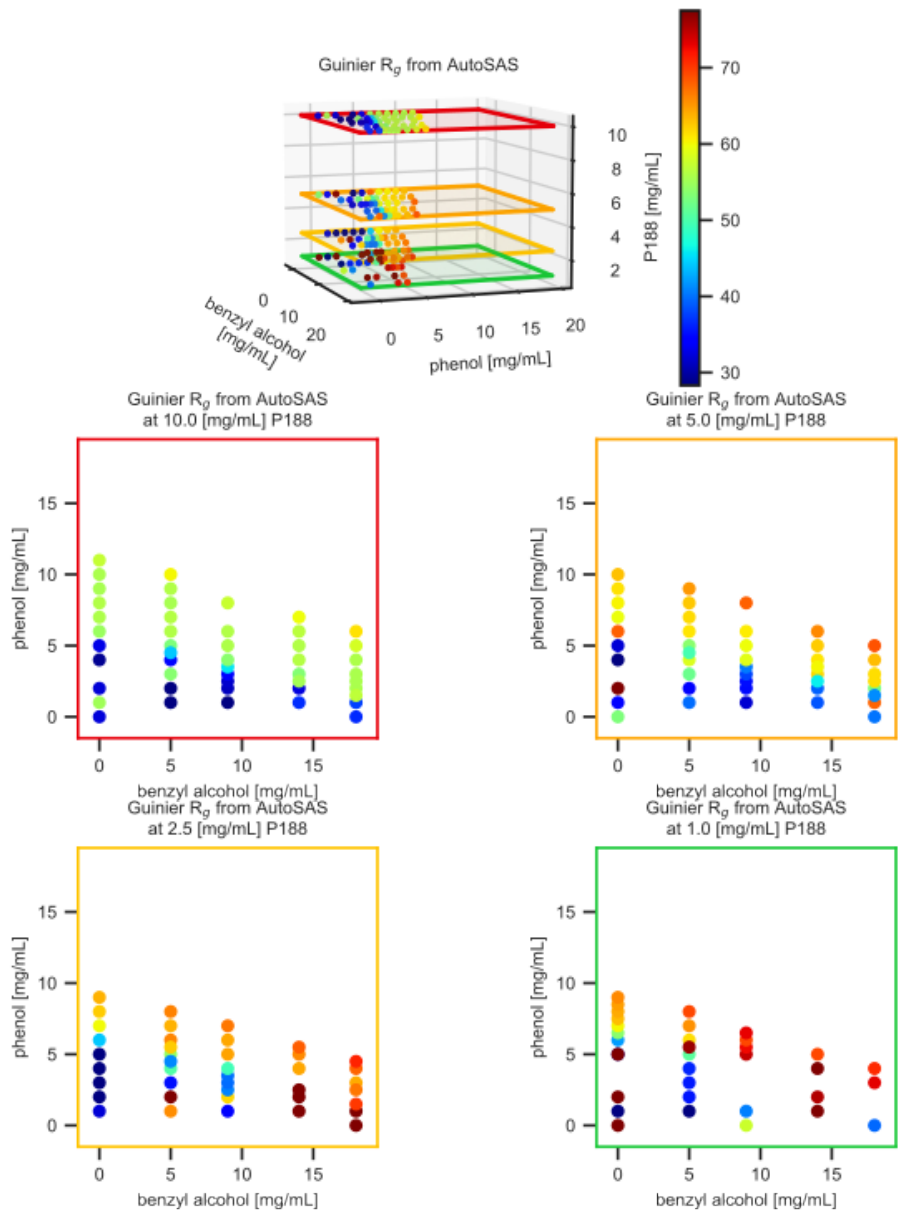

Figure S2.1: Radii of gyration from the Guinier model fit with AutoSAS. 2D cutouts show the fixed P188 compositions and the trends as the phenol and benzyl alcohol vary from 1 mg / mL to 10 mg / mL.

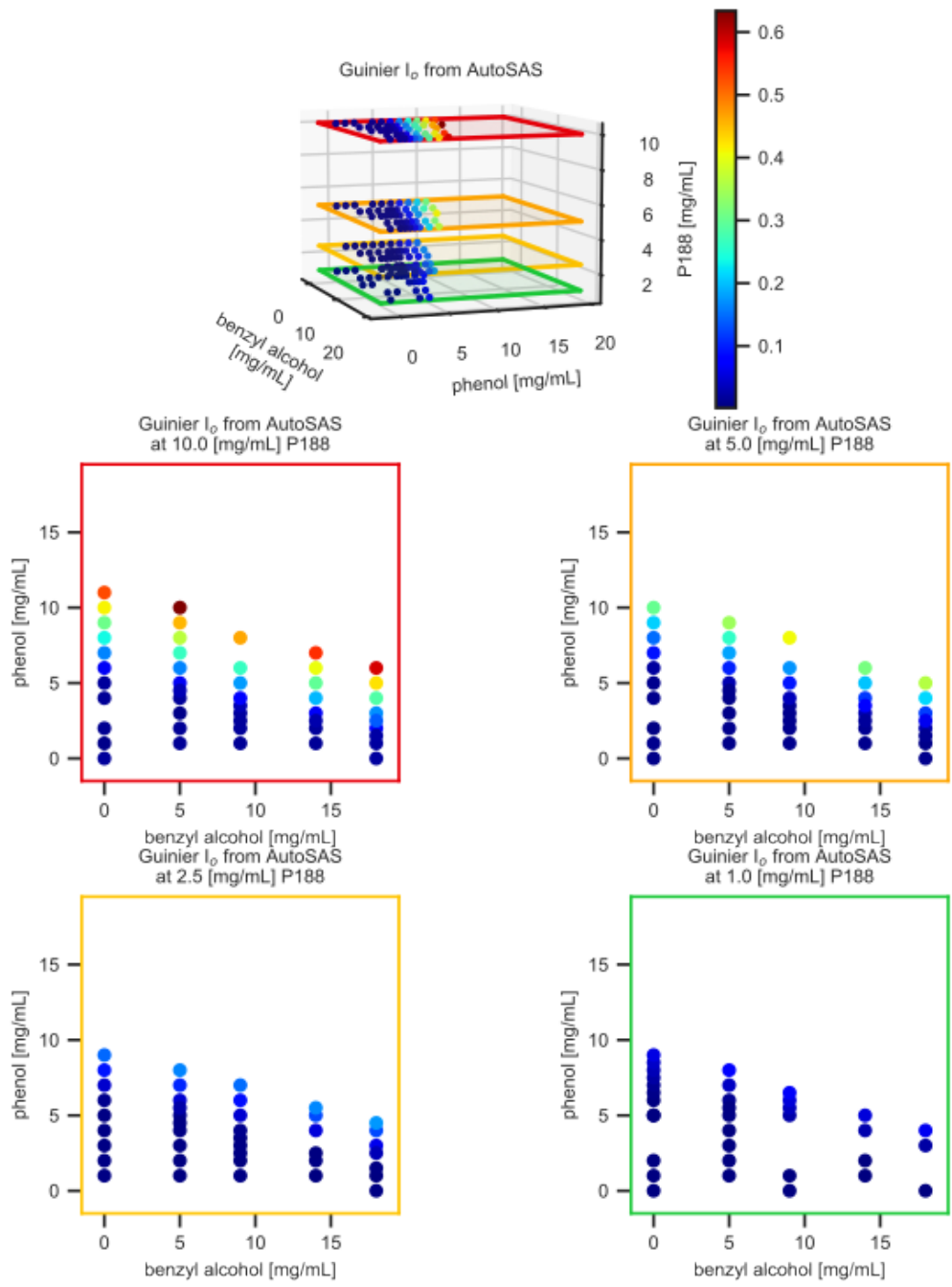

Figure S2.2: $I_0$ parameter from the Guinier model fit with AutoSAS. 2D cutouts show the fixed P188 compositions and the trends as the phenol and benzyl alcohol vary from 1 mg / mL to 10 mg / mL.

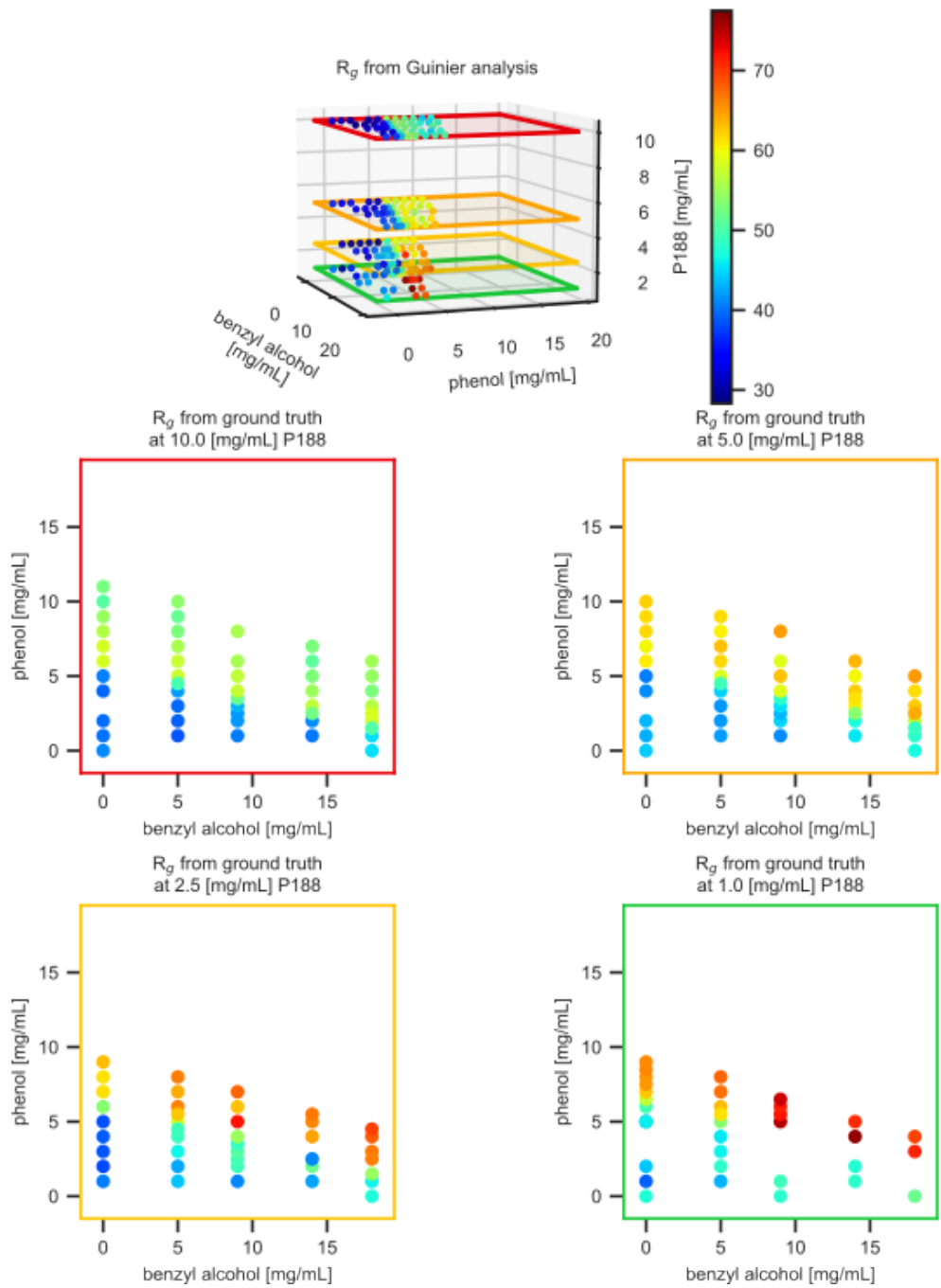

Figure S2.3: Radii of gyration from the Guinier model fit by an expert. 2D cutouts show the fixed P188 compositions and the trends as the phenol and benzyl alcohol vary from 1 mg / mL to 10 mg / mL.

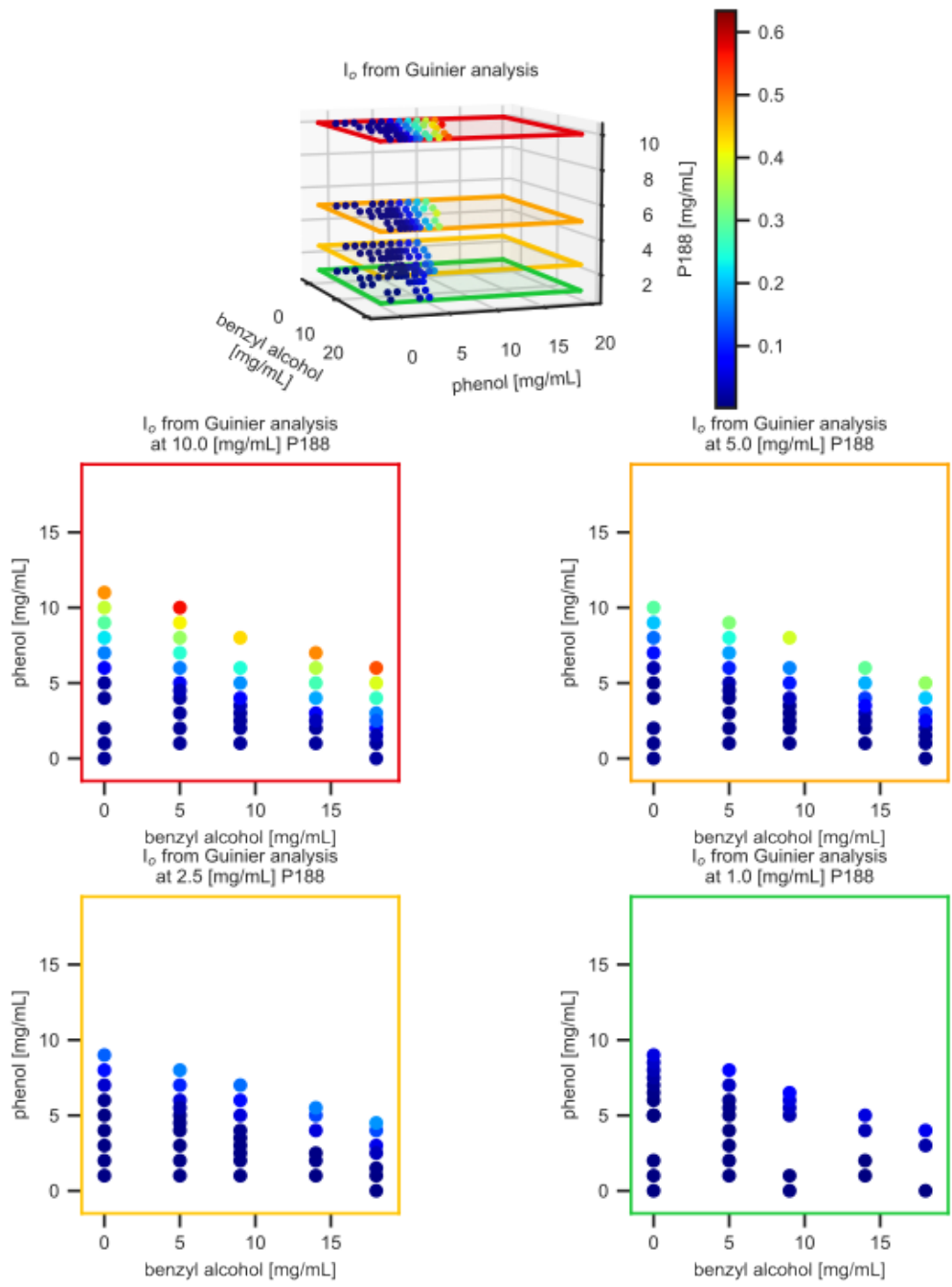

Figure S2.4: Io parameter from the Guinier model fit by an expert. 2D cutouts show the fixed P188 compositions and the trends as the phenol and benzyl alcohol vary from 1 mg / mL to 10 mg / mL.

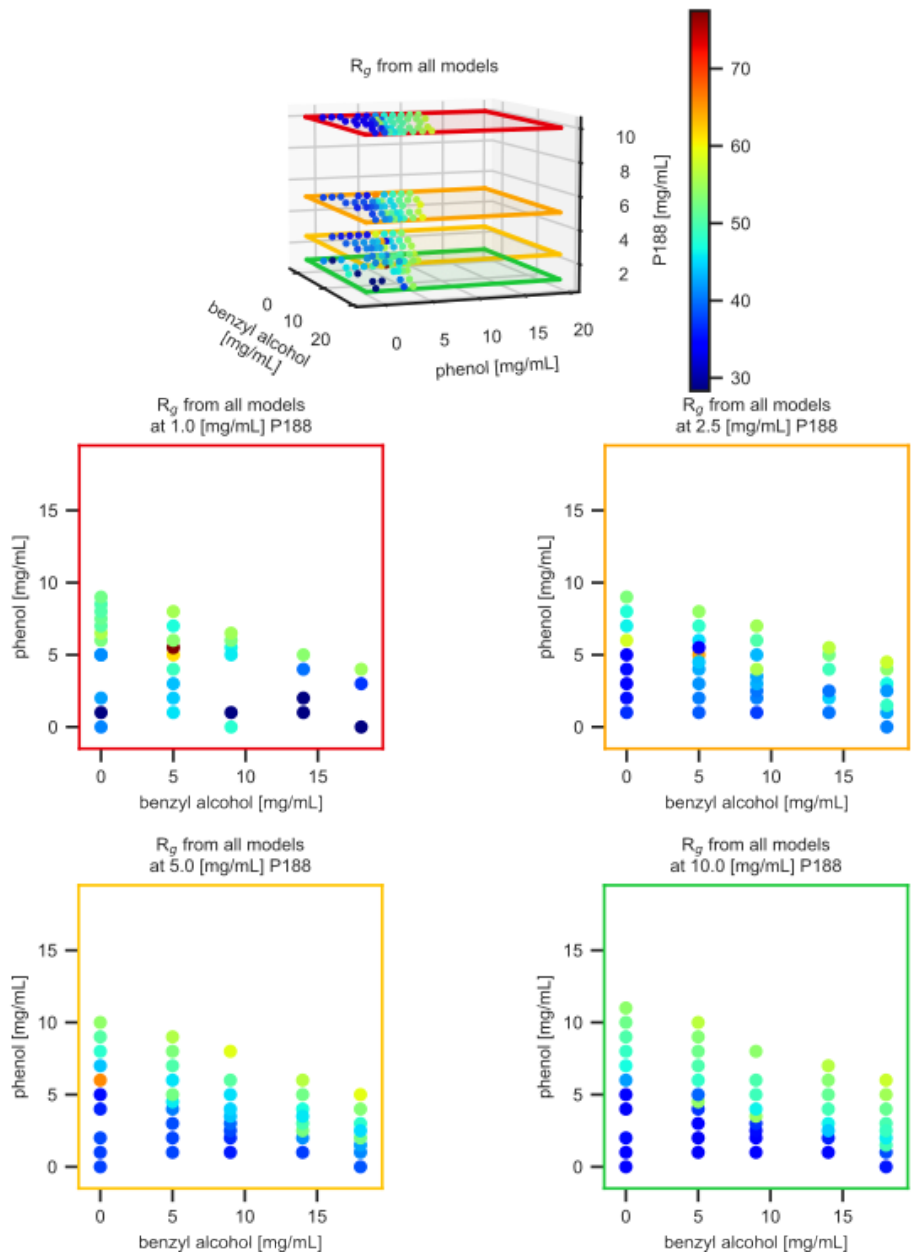

Figure S2.5: Effective radii of gyration for the Occam's razor selected models fit with AutoSAS from 1 mg P188 / mL to 10 mg P188 / mL. The sphere radii are converted to an effective $R_g$ by equation 5 and the mixed model averages the $R_g$ for each of the composite models weighted by their volume fraction contributions.

$$R_g = \sqrt{\frac{5}{3}R^2}$$

(5)

**S3. Phase identification facets**

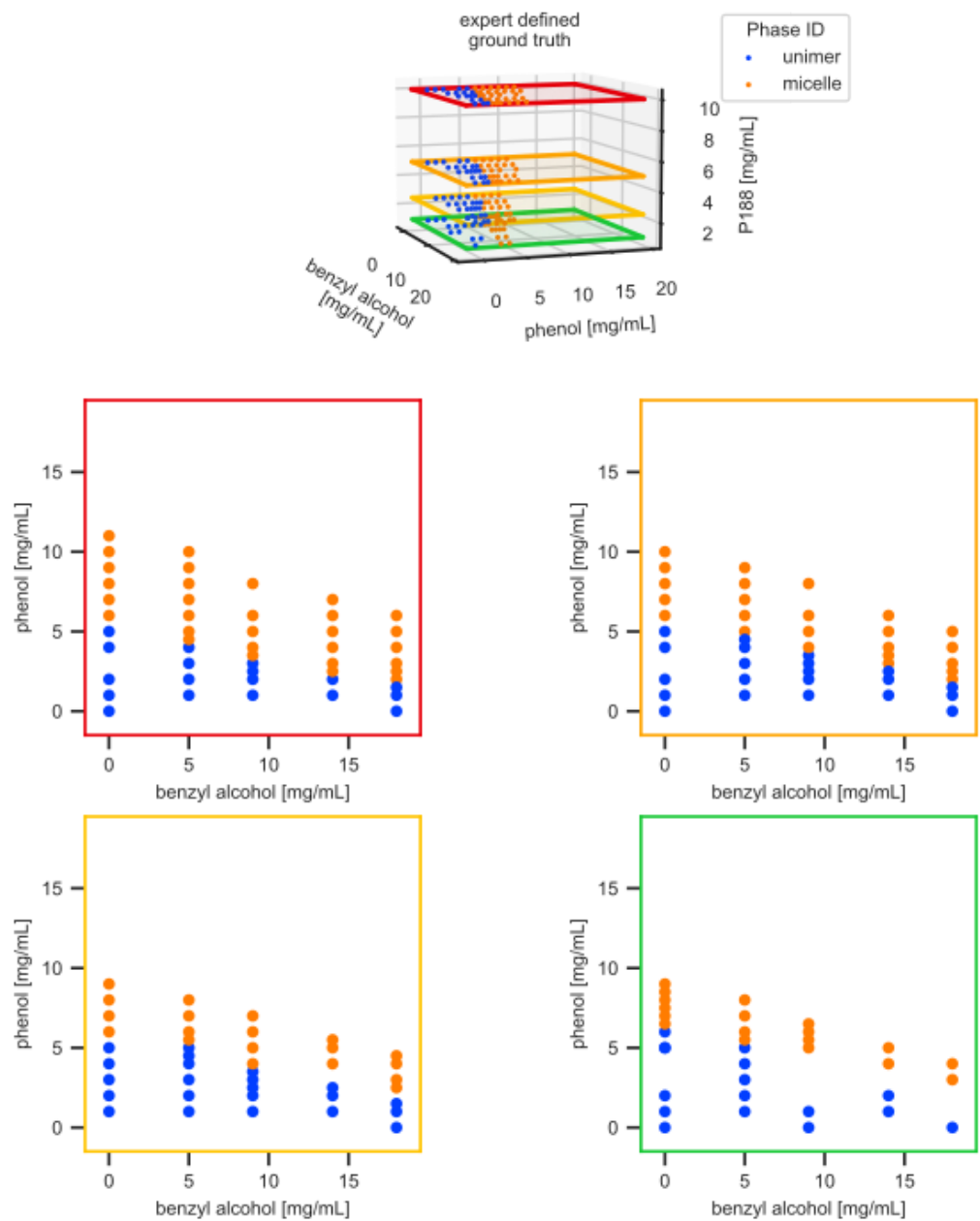

Figure S3.1: Ground truth labels determined by Guinier analysis as a function of P188 concentration (red = 10.0 mg P188 / mL, orange = 5.0 mg P188 / mL, yellow = 2.5 mg P188 / mL, green = 1.0 mg P188 / mL) identified where the $I_o$ parameter abruptly change as the amount of phenol and benzyl alcohol and P188 are increased, in accordance to prior work.[3] Orange markers indicate the presence of micelles and blue markers indicate unimers.

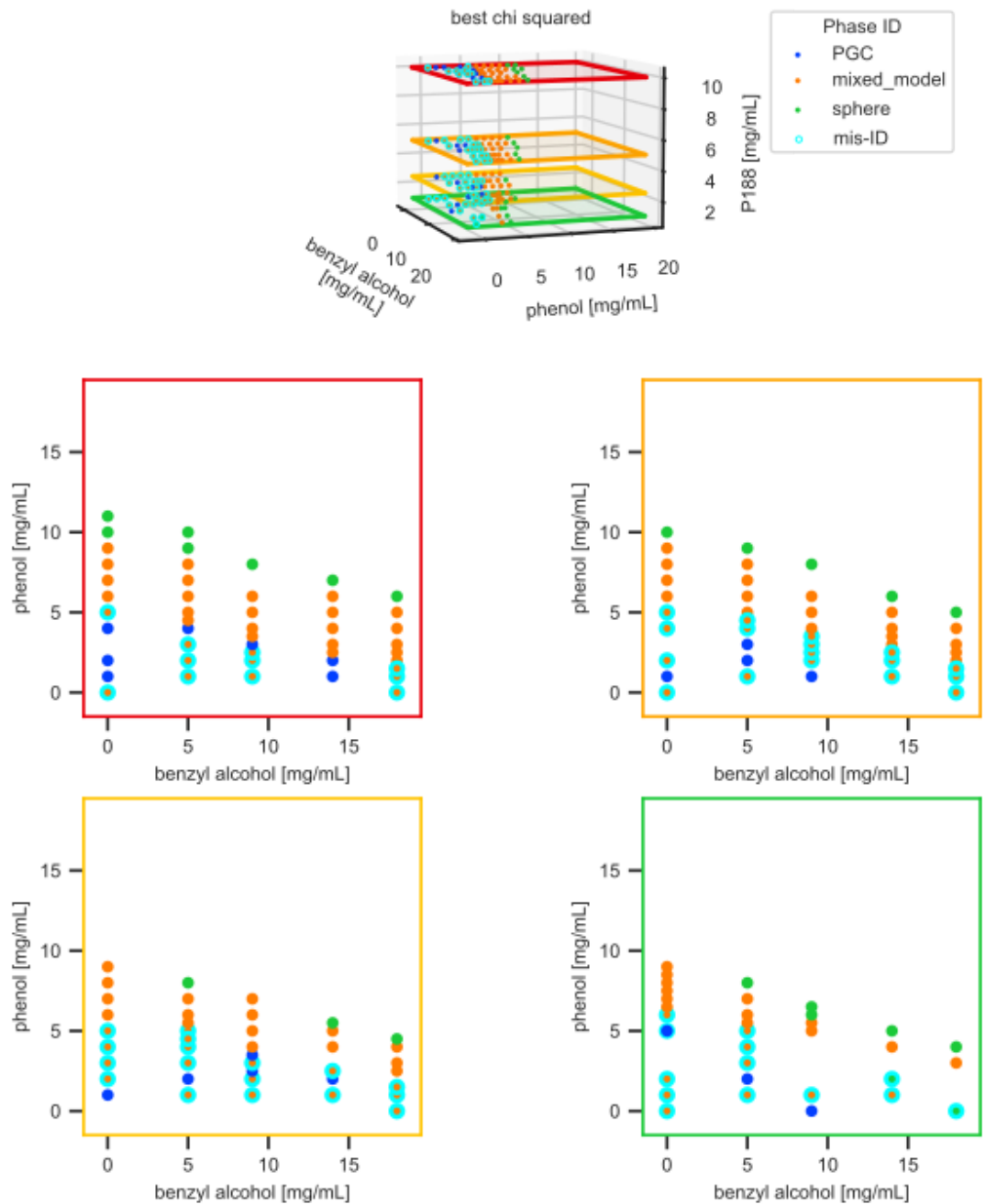

Figure S3.2: AutoSAS model classification by method of best $\chi^2$ for the three contending models as a function of the solution composition (red = 10.0 mg P188 / mL, orange = 5.0 mg P188 / mL, yellow = 2.5 mg P188 / mL, green = 1.0 mg P188 / mL). Color of the marker indicates the classification and mis-identified models are denoted by a cyan border determined by the ground truth in Figure S2.1.

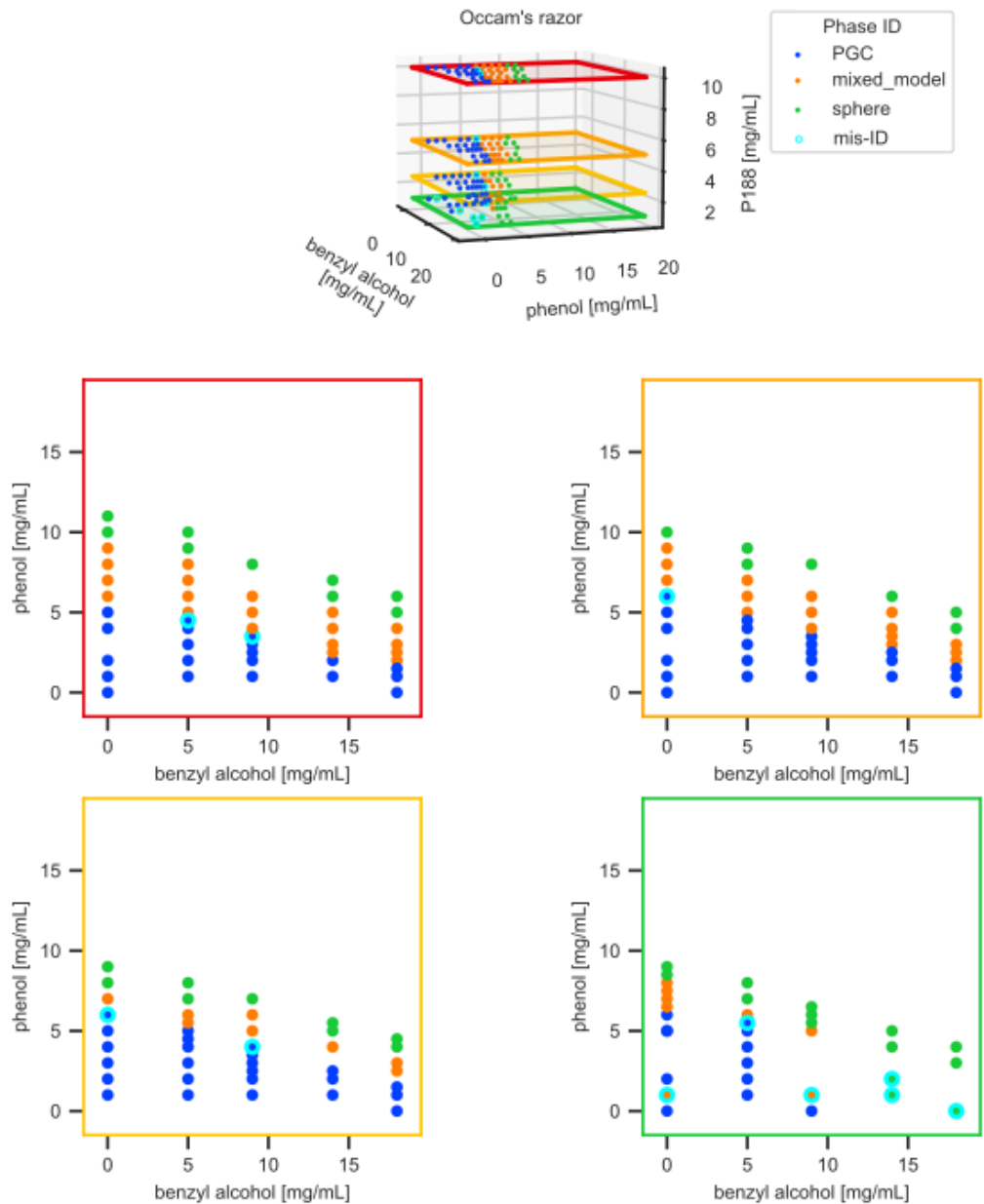

Figure S3.3: AutoSAS model classification by method of Occam's Razor for the three contending models as a function of the solution composition (red = 10.0 mg P188 / mL, orange = 5.0 mg P188 / mL, yellow = 2.5 mg P188 / mL, green = 1.0 mg P188 / mL). Color of the marker indicates the classification and mis-identified models are demarked by a cyan border determined by the ground truth in Figure S2.1.

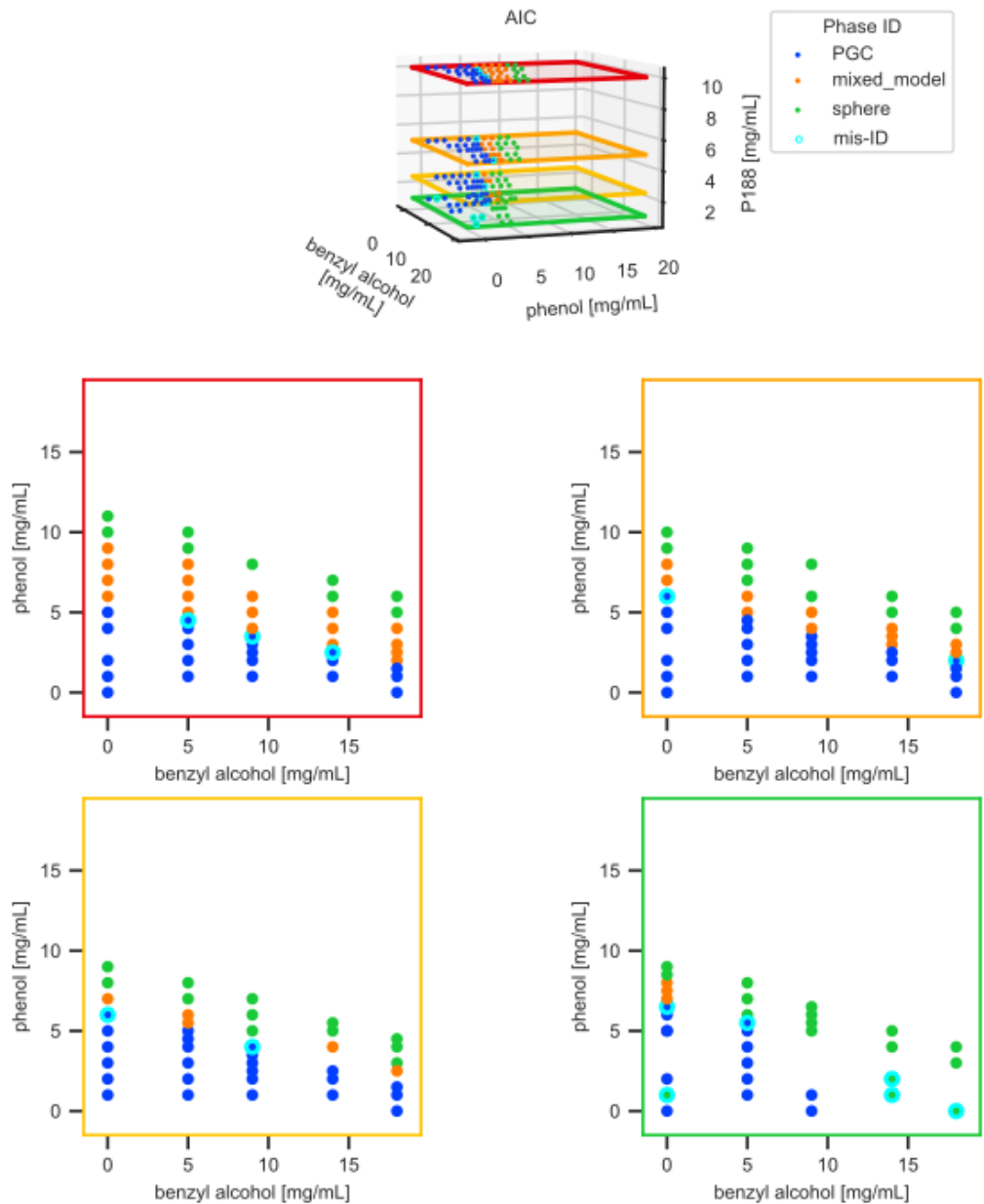

Figure S3.4: AutoSAS model classification by method of lowest AIC for the three contending models as a function of the solution composition (red = 10.0 mg P188 / mL, orange = 5.0 mg P188 / mL, yellow = 2.5 mg P188 / mL, green = 1.0 mg P188 / mL). Color of the marker indicates the classification and mis-identified models are denoted by a cyan border determined by the ground truth in Figure S2.1.

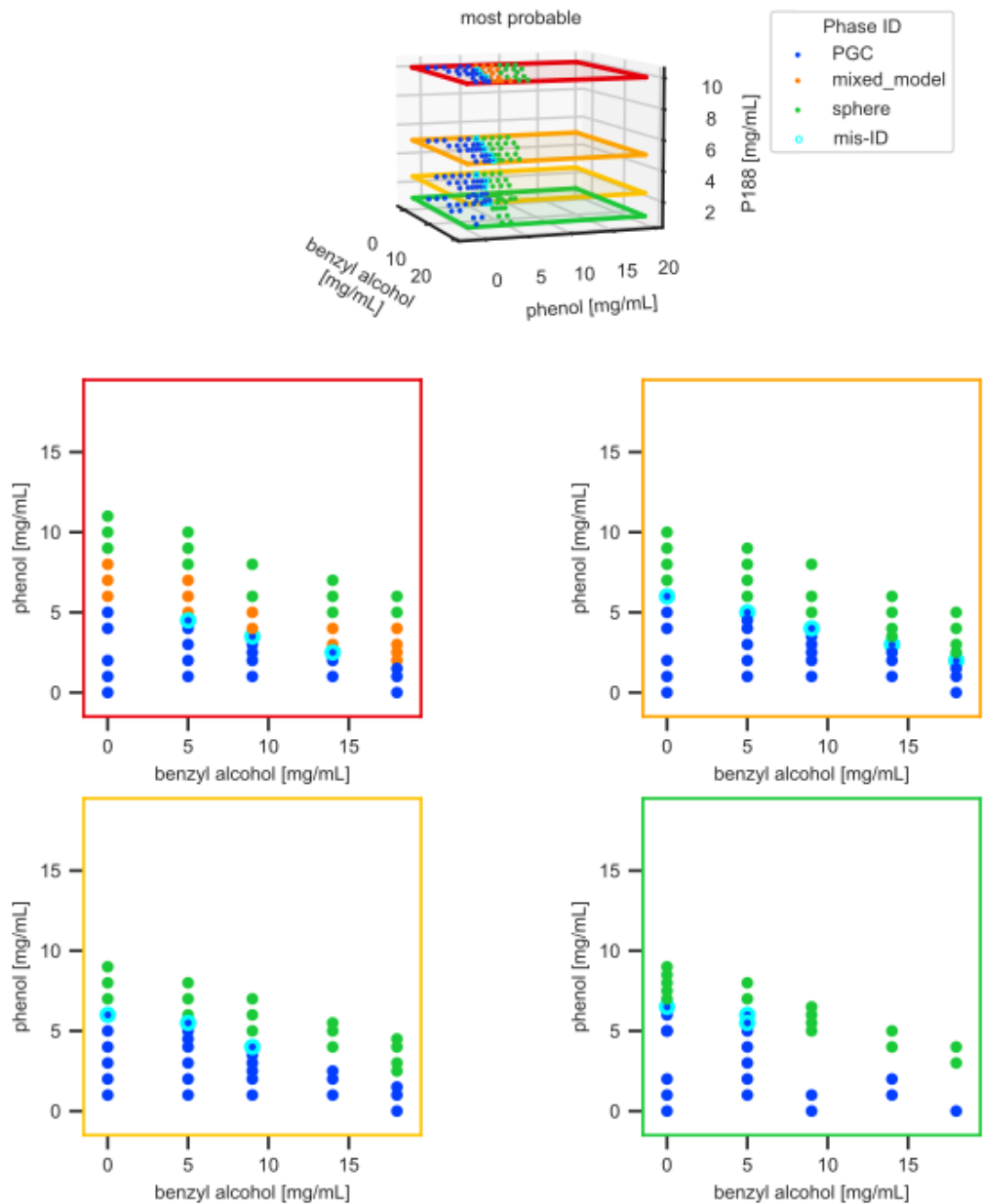

Figure S3.5: AutoSAS model classification by method of most probable for the three contending models as a function of the solution composition (red = 10.0 mg P188 / mL, orange = 5.0 mg P188 / mL, yellow = 2.5 mg P188 / mL, green = 1.0 mg P188 / mL). Color of the marker indicates the classification and mis-identified models are denoted by a cyan border determined by the ground truth in Figure S2.1.

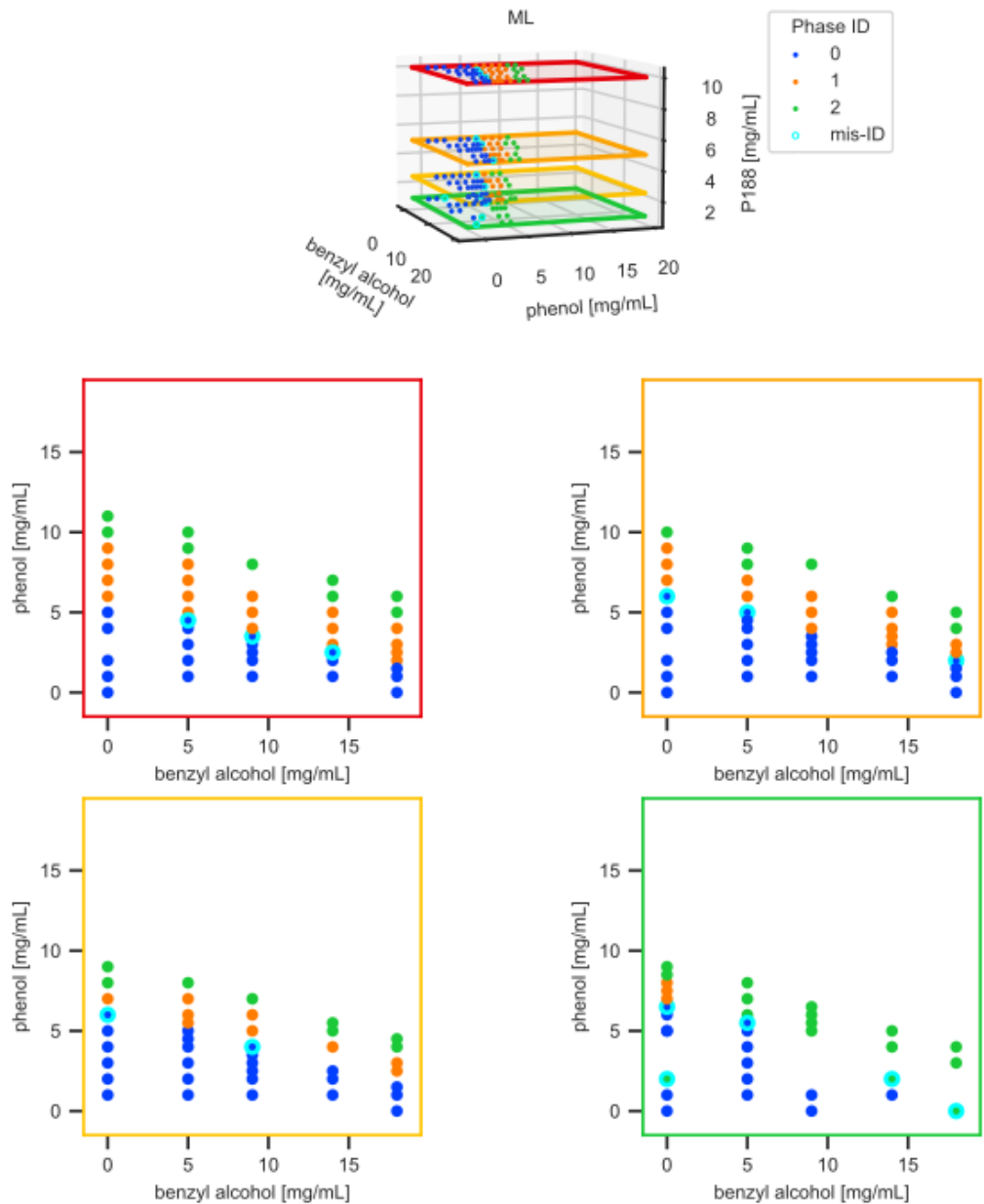

Figure S3.6: The machine learning (ML) classification method as a function of the solution composition (red = 10.0 mg P188 / mL, orange = 5.0 mg P188 / mL, yellow = 2.5 mg P188 / mL, green = 1.0 mg P188 / mL). Color of the marker indicates the classification and mis-identified models are demarked by a cyan border determined by the ground truth in Figure S2.1.

## S4. CMC and MCT boundary transformations

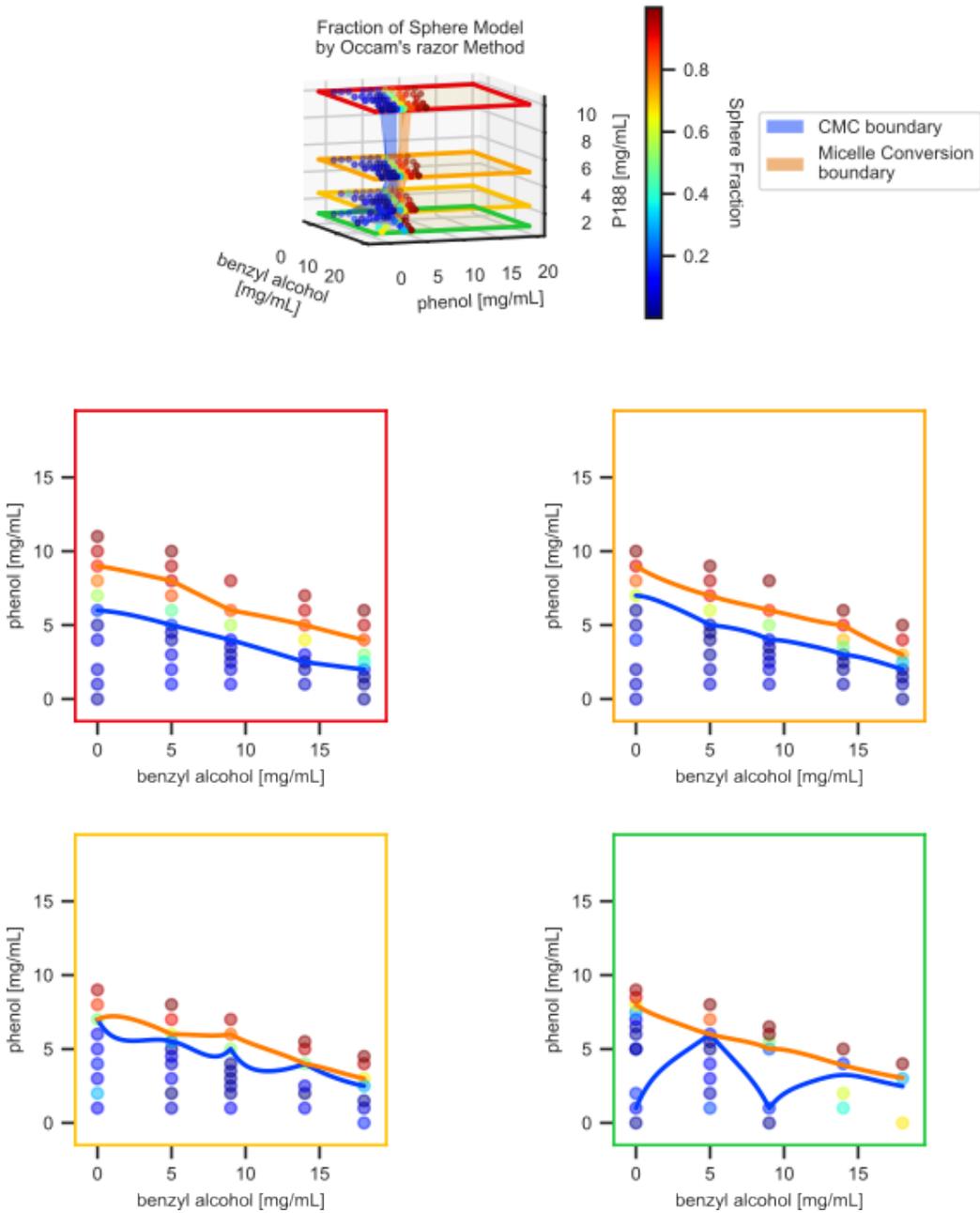

Figure S4.1: The volume fraction of spheres in the solutes as a function of solution composition (red = 10.0 mg P188 / mL, orange = 5.0 mg P188 / mL, yellow = 2.5 mg P188 / mL, green = 1.0 mg P188 / mL). The blue colored surface in (a) and lines in (b)-(c) denote the critical micelle concentration while the orange colored surface and lines denote the micelle conversion threshold determined by the Occam's razor selection criteria, bounding the mixed model domain.

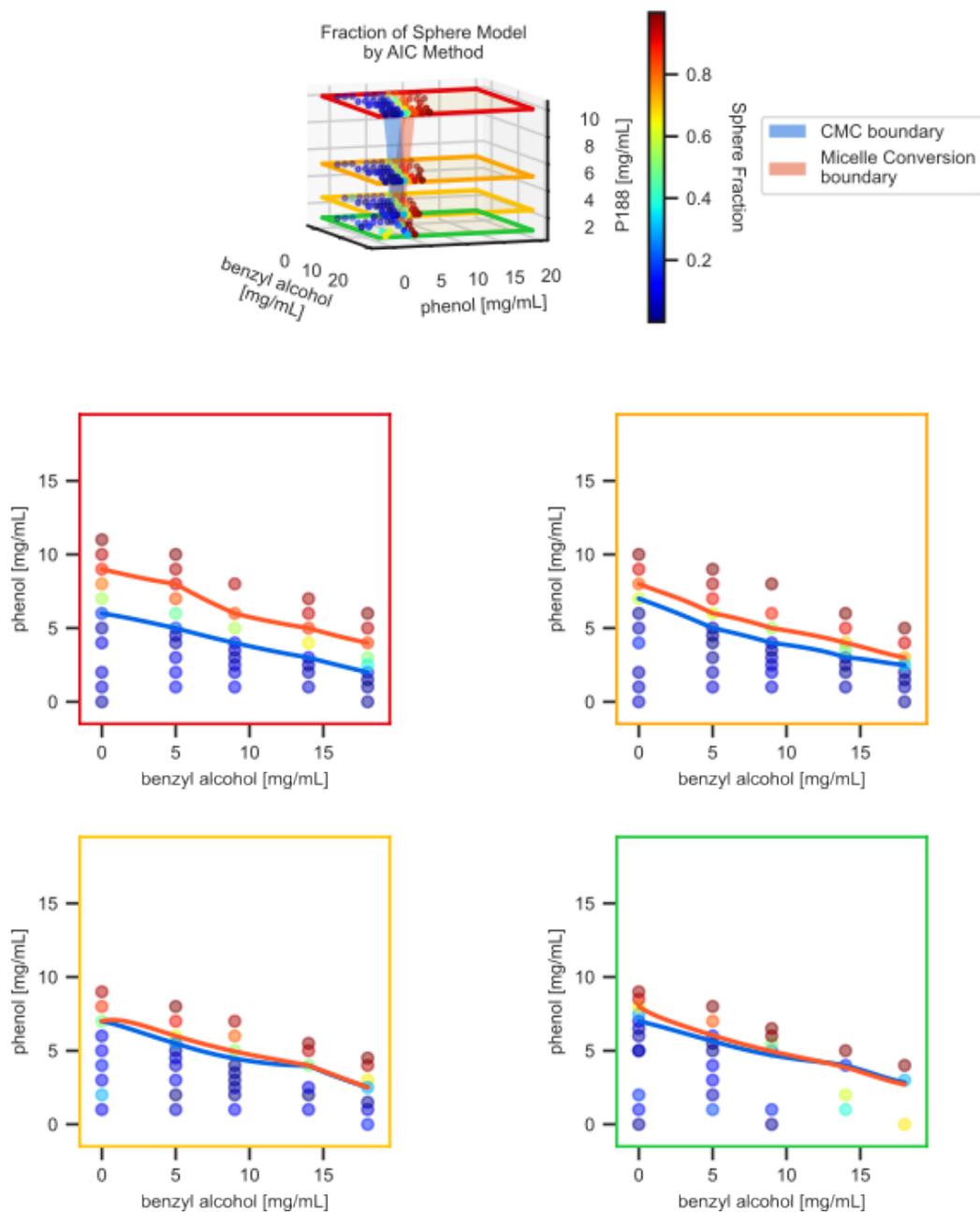

Figure S4.2: The volume fraction of spheres in the solutes as a function of solution composition (red = 10.0 mg P188 / mL, orange = 5.0 mg P188 / mL, yellow = 2.5 mg P188 / mL, green = 1.0 mg P188 / mL). The blue colored surface in (a) and lines in (b)-(c) denote the critical micelle concentration while the orange colored surface and lines denote the micelle conversion threshold determined by the lowest AIC selection criteria, bounding the mixed model domain.

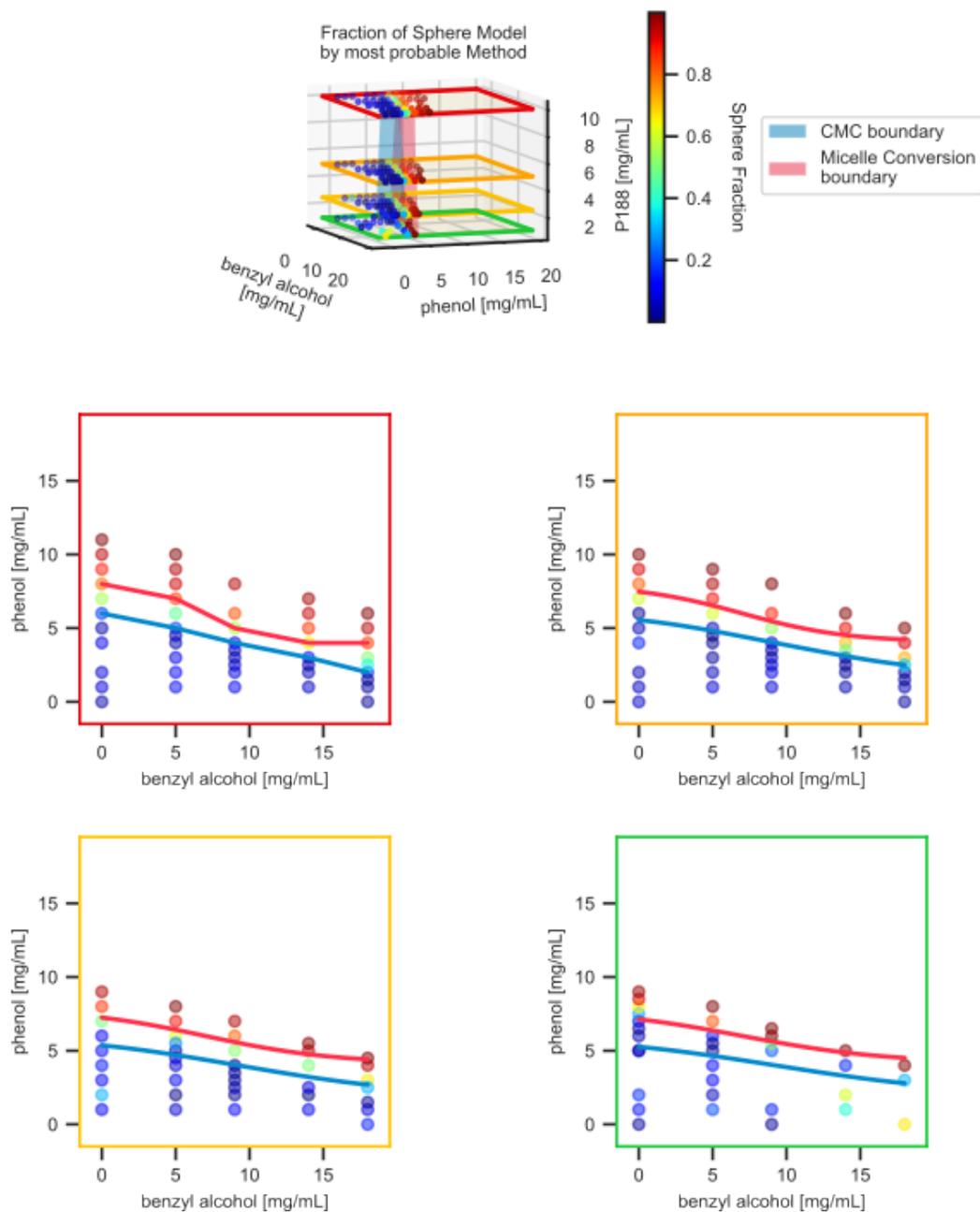

Figure S4.3: The volume fraction of spheres in the solutes as a function of solution composition (red = 10.0 mg P188 / mL, orange = 5.0 mg P188 / mL, yellow = 2.5 mg P188 / mL, green = 1.0 mg P188 / mL). The blue colored surface in (a) and lines in (b)-(c) denote the critical micelle concentration while the orange colored surface and lines denote the micelle conversion threshold determined by the most probable selection criteria, bounding the mixed model domain.

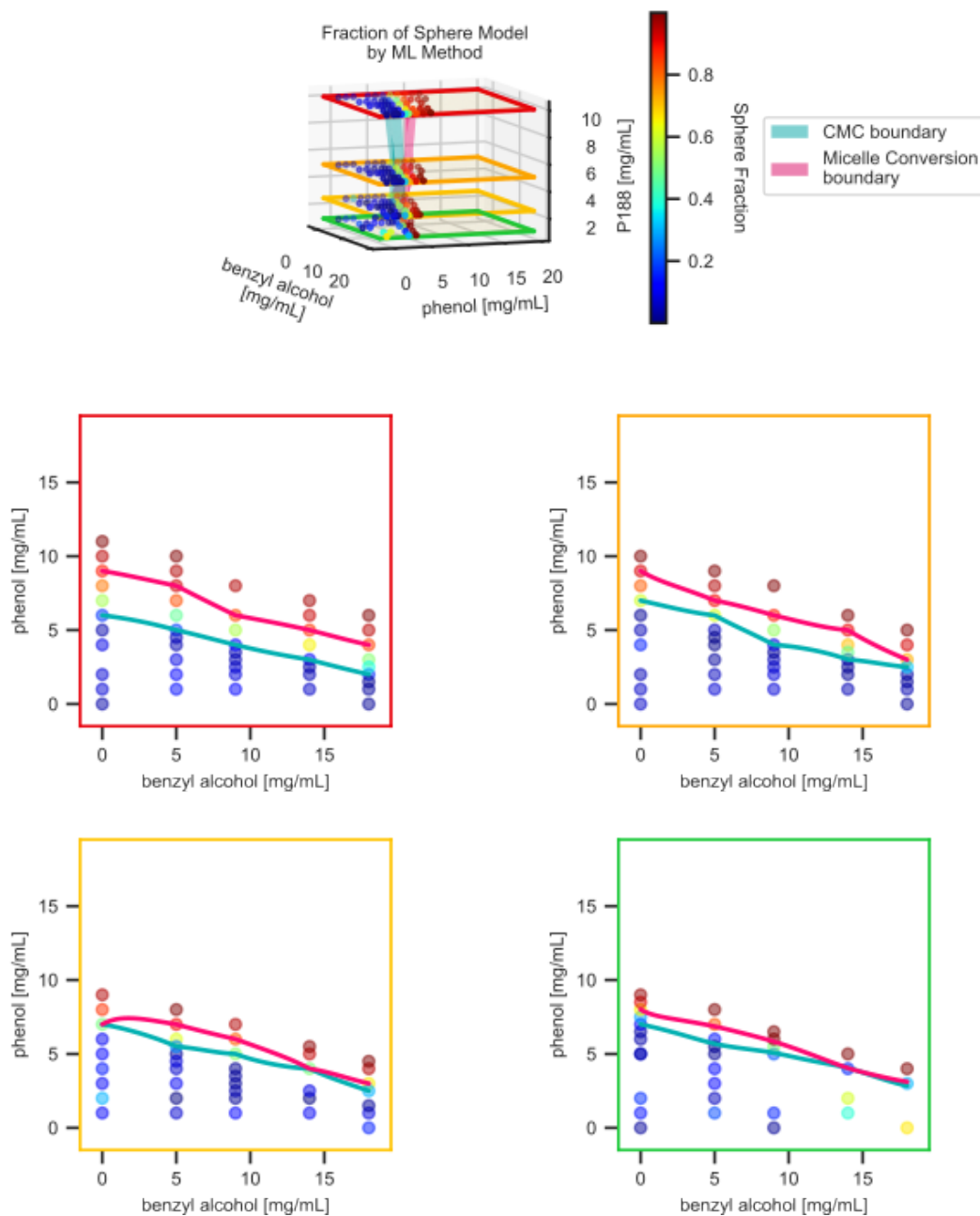

Figures S4.4: The volume fraction of spheres in the solutes as a function of solution composition (red = 10.0 mg P188 / mL, orange = 5.0 mg P188 / mL, yellow = 2.5 mg P188 / mL, green = 1.0 mg P188 / mL). The blue colored surface in (a) and lines in (b)-(c) denote the critical micelle concentration while the orange colored surface and lines denote the micelle conversion threshold determined by the ML classification, bounding the mixed model domain.

**S5. Fitting results for the Guinier analysis**

**S6. Fitting results for the three form factor models**